\documentclass{amsart}
\usepackage{graphicx}
\newtheorem{thm}{Theorem}[section]
\newtheorem{cor}[thm]{Corollary}
\newtheorem{lem}[thm]{Lemma}
\newtheorem{prop}[thm]{Proposition}
\newtheorem{exa}[thm]{Example}
\theoremstyle{definition}
\newtheorem{defn}[thm]{Definition}
\theoremstyle{remark}
\newtheorem{rem}[thm]{Remark}
\numberwithin{equation}{section}
\begin{document}

\title[Orthogonal Trigonometric Polynomials]{Orthogonal Trigonometric Polynomials:
Riemann-Hilbert Analysis and Relations with OPUC}%

\author{Jinyuan Du}%
\address{School of Mathematics and Statistics, Wuhan University, Wuhan 430072, China}%
\email{jydu@whu.edu.cn}%

\author{Zhihua Du }%
\address{Department of Mathematics, Jinan University, Guangzhou 510632, China
}%
\email{zhdu80@126.com}%

%\author{Hua Liu }%
%\address{Department of Mathematics and Physics, Tianjin University of Education and Technology, Tianjin 300222, China
%}%
%\curraddr{Institute of Mathematics, Free University Berlin,
%Arnimallee 3, 14195 Berlin,
%Germany}%
%\email{hualiu@tute.edu.cn}%

\thanks{This work was carried out by the second author (Z. Du) when he was visiting Free University Berlin
from April 11, 2007 to April 10, 2008 on the basis of the State
Scholarship Fund Award of China.}%
\thanks{The authors were supported by the National Natural Science Foundation of
China \#10471107 and RFDP of Higher Education of China
\#20060486001.} %{The third author was supported by the National
%Natural Science Foundation of China \#10601030 and the Scholarship
%of DAAD-K. C. Wong when he was visiting Free University Berlin
%from September 10, 2007 to September 9, 2008.}%
\thanks{The second author appreciates the hospitality from
Institute of Mathematics, Free University Berlin and dearly
expresses his gratitude to Professor H. Begehr for his helpful
support and enlighten discussion. The authors also greatly thank
Professor Hua Liu for his nice ideas and suggestions.
}%

%\date{}%
%\dedicatory{This paper is dedicated to our parents.}%
%\commby{}%
% ----------------------------------------------------------------
\begin{abstract}
In this paper, we study the theory of orthogonal trigonometric
polynomials (OTP). We obtain asymptotics of OTP with positive and
analytic weight functions by Riemann-Hilbert approach and find they
have relations with orthogonal polynomials on the unit circle
(OPUC). By the relations and the theory of OPUC, we also get
four-terms recurrent formulae, Christoffel-Darboux formula and some
properties of zeros for orthogonal trigonometric polynomials.
\end{abstract}
\maketitle
% ----------------------------------------------------------------
\section{Introduction}

The theory of orthogonal polynomials is an important branch of
analysis. It is intimately related to many other mathematical and
physical branches such as continued fractions, special functions,
interpolation, quadrature theory, difference equations, differential
equations and integral equations, etc. in mathematics and quantum
mechanics and mathematical physics including Schr\"odinger
equations, Painlev\'e equations, soliton, scattering and inverse
scattering theory, etc. in physics. In addition, it also has close
connections with combinatorial mathematics and mathematical
statistics. Recently, it is extensively applied to random theory
mainly including random matrices \cite{dft}, random walk, random
permutation and so on.

Because of intimate connections to quantum mechanics and
mathematical physics, the theory of orthogonal polynomials is
abundantly developed since the classic literature \cite{sze} due to
Szeg\"o published for the first time in 1939. In recent years, the
investigations have made great progress in research fields and
research methods. It is notable that new research fields are
ceaselessly extending and new research methods are also continuously
appearing. Among others, Riemann-Hilbert analysis for orthogonal
polynomials is a nice example which is on the basis of a
characterization for orthogonal polynomials in terms of the
classical theory of boundary value problems for analytic functions
\cite{gak,lu,mus}, mainly Riemann-Hilbert problems of matrix form
combined with the so-called steepest descent method for oscillatory
Riemann-Hilbert problems introduced by Deift and Zhou in \cite{dz1}
and further developed in \cite{dz2,dvz}. Now Riemann-Hilbert
approach is a powerful tool for many problems in mathematics and
physics.

Historically, the theory of orthogonal polynomials originated from
the theory of continued fraction. In fact, it has the roots in
attempts to understand and extend the monumental papers of Stieltjes
\cite{stiel}. Continued fractions can be seen as a starting point
for the theory of orthogonal polynomials although the theory of
continued fractions \cite{khin,khov,wall} seemly has fallen out of
favor in recent times. This relationship is of great theoretical and
practical importance. In general, the theory of orthogonal
polynomials can be roughly divided into two main parts, i.e.,
algebraic and analytic aspects which have many things in common.
Just as the names imply, the former has close relations with algebra
while the latter are analytical in methods and questions for
orthogonal polynomials. Usually, in the analytic theory, general
properties are only a smaller part and the greater part is two main
and extensively rich branches, i.e., orthogonal polynomials on the
real line and on the circle. The richness is due to some special
features of the real line and the circle which are not possessed by
other orthogonality regions in the complex plane. Classical real
orthogonal polynomials can be traced back to the 18th century but
their rapid development appeared in the 19th and early 20th century.
However, orthogonal polynomials on the unit circle (simply, OPUC)
are relatively rather younger. Their existence is largely due to
Szeg\"o and Geronimus in the first half of the 20th century. A
recently eminent monograph \cite{sim2,sim3} due to Simon
systematically summarizes and greatly extends what have happened
since then. An excellent survey on orthogonal polynomials of one
(real or complex) variable except for OPUC is given by Totik
\cite{tot} and another good one \cite{gt} by Golinskii and Totik is
partly on orthogonal polynomials on the unit circle.

The investigations of orthogonal polynomials including on the real
line and the unit circle as well general orthogonal polynomials
\cite{st} are truly impressive and the literatures for them are
greatly abundant. However, orthogonal trigonometric polynomials have
been little investigated although this class of orthogonal
polynomials are of theoretical and practical importance and play an
important role in some problems of approximation theory and others
such as quadrature and interpolation \cite{djy1,dhj} and so on. It
is the purpose of this paper to obtain some results in this
direction.

This is the plan of the present paper.

In Section 2 some definitions, notations and preliminaries are
introduced including orthogonal trigonometric polynomials,
orthogonal polynomials on the unit circle, orthogonal Laurent
polynomials on the unit circle and Riemann-Hilbert problems.

The Section 3 and 4 are devoted to Riemann-Hilbert analysis for
orthogonal trigonometric polynomials with positive and analytic
weight functions to obtain their asymptotics by Riemann-Hilbert
approach. More precisely, the theme of Section 3 is the
characterization of orthogonal trigonometric polynomials in terms of
a matrix valued Riemann-Hilbert boundary value problem and the
steepest descent analysis for the characterization by the steepest
descent method originally introduced by Deift and Zhou. However,
asymptotic analysis for orthogonal trigonometric polynomials is the
subject of Section 4, which is on the basis of Section 3.

The main result of Section 5 is that orthogonal trigonometric
polynomials and orthogonal polynomials on the unit circle can be
mutually represented, which makes them relate to each other in
future. As simple consequences, some recurrent formulae,
Christoffel-Darboux formula and some properties of zeros are
obtained for orthogonal trigonometric polynomials.

\section{Some Definitions, Notations and Preliminaries}

Let $\mathbb{D}$ be the open unit disc in the complex plane
$\mathbb{C}$ and $\mu$ be a non-trivial (i.e., with infinite
support) probability measure on the unit circle $\partial
\mathbb{D}$. Throughout, by decomposition, we will write

\begin{equation}
d\mu(\tau)=w(\tau)\frac{d\tau}{2\pi i\tau}+d\mu_{s}(\tau),
\end{equation}
where $\tau\in\partial \mathbb{D}$, $d\mu_{s}$ is the singular part
of $d\mu$ and $w(\tau)=2\pi i\tau d\mu_{ac}/d\tau$ is called a
weight function on the unit circle as $d\mu_{s}\equiv 0$ on
$\partial \mathbb{D}$.

Introduce two kinds of inner product as follows.

\begin{equation}
\langle f,g\rangle_{\mathbb{R}}=\int_{\partial \mathbb{D}}
f(\tau)g(\tau)d\mu(\tau)
\end{equation}
with norm $||f||_{\mathbb{R}}=\big[\int_{\partial
\mathbb{D}}|f(\tau)|^{2}d\mu(\tau)\big]^{1/2}$ for all real valued
functions $f,g$ defined on $\partial \mathbb{D}$ and
\begin{equation}
\langle f,g\rangle_{\mathbb{C}}=\int_{\partial \mathbb{D}}
\overline{f(\tau)}g(\tau)d\mu(\tau)
\end{equation}
with norm $||f||_{\mathbb{C}}=\big[\int_{\partial
\mathbb{D}}|f(\tau)|^{2}d\mu(\tau)\big]^{1/2}$ for all complex
valued functions $f,g$ defined on $\partial \mathbb{D}$. Obviously,
when $f,g$ are real, the above inner products are all the same. In
what follows, we analogously adopt Simon's notations in
\cite{sim2,sim3} with some modifications.

\subsection{Orthogonal Trigonometric Polynomials}

First, we consider the real inner product (2.2).  Respectively, let
$H_{1}$ and $H_{2}$ be real spans of the following over $\mathbb{R}$
linearly independent ordered sets
  \begin{equation}
\Big\{1, \frac {z-z^{-1}}{2i}, \frac {z+z^{-1}}{2},  \ldots,
  \frac{z^{n}-z^{-n}}{2i}, \frac{z^{n}+z^{-n}}{2},  \ldots  \Big\}
  \end{equation}
and
\begin{equation}
\Big\{1, \frac {z+z^{-1}}{2}, \frac {z-z^{-1}}{2i}, \ldots,
  \frac{z^{n}+z^{-n}}{2}, \frac{z^{n}-z^{-n}}{2i}, \ldots  \Big\},
  \end{equation}
$H_{1}^{(n)}$ and $H_{2}^{(n)}$ denote real spans of the first $n+1$
elements of (2.4) and (2.5) and $P_{1}^{(n)}$ and $P_{2}^{(n)}$ be
the orthogonal
 projections on $H^{(n)}_{1}$ and $H^{(n)}_{2}$. Obviously,  while $z=e^{i\theta}$, (2.4) and (2.5)
 become the over $\mathbb{R}$ linearly independent
 ordered trigonometric systems as follows
\begin{equation}
\{\,1,\sin\theta,\cos\theta,\ldots,\sin{n\theta},\cos{n\theta},\ldots\,\}
 \end{equation}
 and
\begin{equation}
\{\,1,\cos\theta,\sin\theta,\ldots,\cos{n\theta},\sin{n\theta},\ldots\,\}.
 \end{equation}

 Set $\mu_{0}^{(0)}(z)=u_{0}^{(0)}(z)=1$, and
\begin{equation}
\mu_{n}^{(0)}(z)=\left\{
\begin{array}{lllllll}
\displaystyle \frac{z^{k}+z^{-k}}{2},\hspace{3mm} n=2k, \\
&k=1,2,\ldots\\
\displaystyle \frac{z^{k}-z^{-k}}{2i},\hspace{3mm} n=2k-1,
\end{array}
\right.
\end{equation}

 \begin{equation}
u_{n}^{(0)}(z)=\left\{
\begin{array}{lllllll}
\displaystyle \frac{z^{k}-z^{-k}}{2i},\hspace{3mm} n=2k, \\
&k=1,2,\ldots\\
\displaystyle \frac{z^{k}+z^{-k}}{2},\hspace{3mm} n=2k-1
\end{array}
\right.
\end{equation} as well as
\begin{equation}
\mu_n(z)=\frac{(\mathbf{1}-P_{1}^{(n-1)})\mu_{n}^{(0)}(z)}{\|(\mathbf{1}-P_{1}^{(n-1)})\mu_{n}^{(0)}\|_{\mathbb{R}}},
\end{equation}

\begin{equation}
u_n(z)=\frac{(\mathbf{1}-P_{2}^{(n-1)})u_{n}^{(0)}(z)}{\|(\mathbf{1}-P_{2}^{(n-1)})u_{n}^{(0)}\|_{\mathbb{R}}},
\end{equation} where $\mathbf{1}$ is the identity operator, then
$\{\mu_{n}(e^{i\theta})\}$ and $\{u_{n}(e^{i\theta})\}$ are
respectively two unique systems of orthonormal trigonometric
polynomials (simply, OTP) on the unit circle with respect to $\mu$
according to the orders of the above systems (2.6) and (2.7), i.e.,
from left to right. By continuation of argument, for $z\in
\mathbb{C}\backslash \{0\}$, we call $\{\mu_{n}(z)\}$ and
$\{u_{n}(z)\}$ the first and second systems of orthonormal Laurent
polynomials on the unit circle with respect to $\mu$. More
precisely, we call $\mu_{n}(z)$ and $u_{n}(z)$ the first and second
orthonormal Laurent polynomials of order $n$ on the unit circle with
respect to $\mu$. And $\mu_{n}(e^{i\theta})$ and
$u_{n}(e^{i\theta})$ are correspondingly called the first and second
orthonormal trigonometric polynomials of order $[\frac{n-1}{2}]+1$
on the unit circle with respect to $\mu$. Here $[x]$ denotes the
greatest integer less than or equal to $x$. It must be noted that
there exist two orthonormal trigonometric polynomials of the same
order which are orthogonal to each other. So it happens since the
above orthogonalization process is ordered. In \cite{dg}, Du and Guo
discussed the second orthomormal Laurent and trigonometric
polynomials on the unit circle and obtained six-terms recurrent
formulae and the simplicity and other properties of their zeros
which were also obtained in \cite{djy1}.

For convenience, in what follows, we always write
\begin{equation}
\pi_{k}(z)=\mu_{2k-1}(z),\,\,\,\,\sigma_{k}(z)=\mu_{2k}(z)
\end{equation}
and
\begin{equation}
\varrho_{k}(z)=u_{2k-1}(z),\,\,\,\,\rho_{k}(z)=u_{2k}(z),
\end{equation}
where $k=1,2,\ldots$\,. Therefore,
\begin{equation}
\langle\pi_{k},\sigma_{k}\rangle_{\mathbb{R}}=0,\,\,\langle\varrho_{k},\rho_{k}\rangle_{\mathbb{R}}=0,\,\,\,k=1,2,\ldots
\end{equation}
and
\begin{equation}
a_{k}\sigma_{k}(z)=\mu^{(0)}_{2k}(z)-\beta_{k}b_{k}\pi_{k}(z)+L_{1,k}(z),
\,\,\,\,L_{1,k}\in H^{(2k-2)}_{1},
\end{equation}
\begin{equation}
c_{k}\rho_{k}(z)=u^{(0)}_{2k}(z)-\gamma_{k}d_{k}\varrho_{k}(z)+L_{2,k}(z),
\,\,\,\,L_{2,k}\in H^{(2k-2)}_{2},
\end{equation}
where
\begin{equation}
a_{k}=\|(\mathbf{1}-P_{1}^{(2k-1)})\mu_{2k}^{(0)}\|_{\mathbb{R}},
\,b_{k}=\|(\mathbf{1}-P_{1}^{(2k-2)})\mu_{2k-1}^{(0)}\|_{\mathbb{R}},
\,\beta_{k}=b_{k}^{-1}\langle
\mu_{2k}^{(0)},\pi_{k}\rangle_{\mathbb{R}}
\end{equation}
and
\begin{equation}
c_{k}=\|(\mathbf{1}-P_{2}^{(2k-1)})u_{2k}^{(0)}\|_{\mathbb{R}},
\,d_{k}=\|(\mathbf{1}-P_{2}^{(2k-2)})u_{2k-1}^{(0)}\|_{\mathbb{R}},
\,\gamma_{k}=d_{k}^{-1}\langle
u_{2k}^{(0)},\varrho_{k}\rangle_{\mathbb{R}}.
\end{equation}
Throughout, as a convention, it is convenient to take
$\sigma_{0}=1$, $\pi_{0}=0$ and $\beta_{0}=0$ as well as
$a_{0}=b_{0}=1$. In addition, it is obvious that
\begin{equation}
\langle\mu_{l}^{(0)},\sigma_{k}\rangle_{\mathbb{R}}=\langle\mu_{l}^{(0)},\pi_{k}\rangle_{\mathbb{R}}=0,\,\,l=0,1,\ldots,2k-2
\end{equation}
and
\begin{equation}
\langle u_{l}^{(0)},\rho_{k}\rangle_{\mathbb{R}}=\langle
u_{l}^{(0)},\varrho_{k}\rangle_{\mathbb{R}}=0,\,\,l=0,1,\ldots,2k-2.
\end{equation}

Identifying the unit circle with the interval $[0,\,2\pi)$ via the
map $\theta\mapsto e^{i\theta}$, all the same, we write
$\sigma_{k}(\theta)=\sigma_{k}(e^{i\theta})$,
$\pi_{k}(\theta)=\pi_{k}(e^{i\theta})$,
$\rho_{k}(\theta)=\rho_{k}(e^{i\theta})$ and
$\varrho_{k}(\theta)=\varrho_{k}(e^{i\theta})$, $k=1,2,\ldots$\,.
Obviously, by (2.8),(2.9),(2.19) and (2.20),
$\sigma_{k},\pi_{k},\rho_{k},\varrho_{k}$ are orthonormal
trigonometric polynomials of order $k$ on the unit circle with
respect to $\mu$ and satisfy (2.14)-(2.18). Thus (2.15) and (2.16)
can be written as
\begin{equation}
a_{k}\sigma_{k}(\theta)=\cos(k\theta)-\beta_{k}b_{k}\pi_{k}(\theta)+T_{1,k-1}(\theta),
\,\,\,\,T_{1,k-1}\in H^{T}_{k-1},
\end{equation}
and
\begin{equation}
c_{k}\rho_{k}(\theta)=\sin(k\theta)-\gamma_{k}d_{k}\varrho_{k}(\theta)+T_{2,k-1}(\theta),
\,\,\,\,T_{2,k-1}\in H^{T}_{k-1},
\end{equation}
where $k=1,2,\ldots$, $H_{k-1}^{T}$ denotes the set of all real
trigonometric polynomials of order at most $k-1$, equivalently, the
span of the first $2k-1$ elements of trigonometric system (2.6) or
(2.7) and $H_{0}^{T}=\mathbb{R}$.

By straightforward calculations, using (2.15) and (2.16) or (2.21)
and (2.22), it is easy to get a basis transform formula between the
first orthonormal Laurent or trigonometric polynomials on the unit
circle and the second ones. The detail is left to the reader.

\subsection{Orthogonal and Orthogonal Laurent Polynomials on the Unit Circle}

Next, we turn to the complex inner product (2.3). Applying
Gram-Schmidt orthogonalization process to the over $\mathbb{C}$
linearly independent system of $\{\,1,z,z^{2},\ldots\}$, we can
obtain the monic orthogonal polynomials $\Phi_{n}(z)$  and the
orthonormal polynomials $\varphi_{n}(z)$ on the unit circle
satisfying
\begin{equation}
\Phi_{n}(z)=z^{n}+\,\,{\rm lower\,\,
order},\,\,\,\,\varphi_{n}(z)=\kappa_{n}z^{n}+\,\,{\rm
lower\,\,order}
\end{equation}
with $\kappa_{n}>0$ and
\begin{equation}
\langle\Phi_{m},\Phi_{n}\rangle_{\mathbb{C}}=\kappa_{n}^{-2}\delta_{mn},\,\,\,\,
\langle\varphi_{m},\varphi_{n}\rangle_{\mathbb{C}}=\delta_{mn}.
\end{equation}
Then $\|\Phi_{n}\|^{2}_{\mathbb{C}}=\kappa^{-2}_{n}$ and
$\varphi_{n}=\Phi_{n}/\|\Phi_{n}\|_{\mathbb{C}}$ follow from (2.23)
and (2.24).

The reversed polynomial $Q^{*}$  is defined by
\begin{equation}
Q^{*}_{n}(z)=z^{n}\overline{Q_{n}(1/\overline{z})}
\end{equation} for any polynomial $Q_{n}$ of order
$n$, i.e.,
\begin{equation}
Q^{*}_{n}(z)=\sum_{j=0}^{n}\overline{q}_{n-j}z^{j}\,\,{\rm
while}\,\,Q_{n}(z)=\sum_{j=0}^{n}q_{j}z^{j}.
\end{equation}

One of the famous properties of orthogonal polynomials on the unit
circle is the Szeg\"o recurrence \cite{sim1,sim2,sim3,sze}. That is
to say that
\begin{equation}
\Phi_{n+1}(z)=z\Phi_{n}(z)-\overline{\alpha}_{n}\Phi^{*}_{n}(z),
\end{equation}
where $\alpha_{n}=-\overline{\Phi_{n+1}(0)}$ are called Verblunsky
coefficients. By convention, $\alpha_{-1}=-1$ (see \cite{sim2}).
Szeg\"o recurrence (2.27) is extremely useful in the theory of OPUC.
By (2.25) and (2.27), we also have
\begin{equation}
\Phi^{*}_{n+1}(z)=\Phi^{*}_{n}(z)-\alpha_{n}z\Phi_{n}(z).
\end{equation}

For the orthonormal polynomials $\varphi_{n}(z)$, the recurrences
become
\begin{equation}
\varphi_{n+1}(z)=\eta_{n}^{-1}(z\varphi_{n}(z)-\overline{\alpha}_{n}\varphi^{*}_{n}(z))
\end{equation}
and
\begin{equation}
\varphi^{*}_{n+1}(z)=\eta_{n}^{-1}(\varphi^{*}_{n}(z)-\alpha_{n}z\varphi_{n}(z))
\end{equation}
with $\eta_{n}=(1-|\alpha_{n}|)^{1/2}=\kappa_{n}/\kappa_{n+1}$.

Making use of (2.29) and (2.30), the well-known Christoffel-Darboux
formula for OPUC is obtained, namely
\begin{align}
\sum_{j=0}^{n}\overline{\varphi_{j}(\zeta)}\varphi_{j}(z)&=\frac{\overline{\varphi^{*}_{n+1}(\zeta)}
\varphi^{*}_{n+1}(z)-\overline{\varphi_{n+1}(\zeta)}
\varphi_{n+1}(z)}{1-\overline{\zeta} z}\\
&=\frac{\overline{\varphi^{*}_{n}(\zeta)}
\varphi^{*}_{n}(z)-z\overline{\zeta}\,\overline{\varphi_{n}(\zeta)}
\varphi_{n}(z)}{1-\overline{\zeta} z}
\end{align}
for any $z,\zeta\in \mathbb{C}$ with $z\overline{\zeta}\neq1$.

Similarly, in \cite{sim2}, using the complex inner product, Simon
defined two classes of orthogonal Laurent polynomials on the unit
circle $\chi_{n}(z)$ and $x_{n}(z)$ by respectively applying
Gram-Schmidt orthogonalization process to two complex linearly
independent ordered sets of $\{1, z, z^{-1},\ldots, z^{n}, z^{-n},
\ldots\}$ and $\{1, z^{-1}, z, \ldots,z^{-n}, z^{n}, \ldots\}$,
which are called the CMV basis and the alternate CMV basis by Simon
and originally discussed in \cite{cmv} and maybe earlier in other
places. They are related to OPUC by
\begin{equation}
\chi_{2n-1}(z)=z^{-n+1}\varphi_{2n-1}(z),\,\,\,\,\chi_{2n}(z)=z^{-n}\varphi^{*}_{2n}(z)
\end{equation}
and
\begin{equation}
x_{2n-1}(z)=z^{-n}\varphi^{*}_{2n-1}(z),\,\,\,\,x_{2n}(z)=z^{-n}\varphi_{2n}(z).
\end{equation}
Moreover, $x_{n}(z)=\overline{\chi_{n}(1/\overline{z})}$. The two
classes of orthogonal Laurent polynomials have closely relation to
CMV matrices which are one of the most interesting developments in
the theory of OPUC discovered by Cantero, Moral, and Vel\'azquez in
their eminent paper \cite{cmv}. Recently, the theory of CMV matrices
are rapidly developed by contributions of many mathematicians
\cite{kn,ls,li,nen,sim4} under the important impact of Simon since
the CMV matrices provided a powerful tool for many mathematical and
physical branches such as perturbation theory of operators, spectral
theory, random matrices and integrable systems and so on.

\subsection{Riemann Boundary Value Problems}

As the final, we sketchily introduce the theory of Riemann boundary
value problems (sometimes, also called Riemann-Hilbert boundary
value problems in some texts). The detailed theory can be found in
\cite{gak,lu,mus}. In the present subsection, we adopt the notations
in \cite{dd}.

Let $L$ be a set of smooth closed curves $L_1,L_2,\ldots,L_n$ in the
complex plane, non-intersecting to each other and positively
oriented for each $L_j$. \,\,Therefore $L$ is also positively
oriented and denoted by $L=\sum\limits_{j=1}^nL_j$. $L$ divides the
extended complex plane into a finite number of regions. The region
containing the point at infinity is put into $S^-$, the region
neighboring to it is put into $S^+$, and so on. Then the entire
complex plane is divided into two open sets $S^+$ and $S^-$.
Certainly, we may put the point at infinity in $S^+$ at the
beginning and then proceed as above, then $S^+$ and $S^-$ are
interchanged in position and all $L_j$  and hence $L$ are oppositely
oriented.

Let $\varphi$ be defined on $L$ and H\"{o}lder continuous, denoted
as $\varphi\in H(L)$. Now we introduce the Cauchy singular integral
operator
\begin{equation}
\mathbf{C}[\varphi](z)=\left\{
\begin{array}{ll}
\displaystyle\frac{1}{2\pi
i}\int_{L}\frac{\varphi(\tau)}{\tau-z}d\tau,\hspace{3mm}
&z\not\in L,\\[6mm]
\displaystyle\frac{1}{\pi
i}\int_{L}\frac{\varphi(\tau)}{\tau-t}d\tau,\hspace{3mm} &z=t\in L,
\end{array}
\right.
\end{equation} and the projection operators
\begin{equation}
\mathbf{C}^{\pm}[\varphi](z)=\left\{
\begin{array}{ll}
\mathbf{C}[\varphi](z),\hspace{3mm}&z\in S^{\pm},\\[3mm]
{\pm}\displaystyle\frac{1}{2}\varphi(t)+\displaystyle\frac{1}{2}\mathbf
{C}[\varphi](t), &z=t\in L.
\end{array}
\right.
\end{equation}

Let $A^+H$($A^-H$) denote the class of functions analytic in
$S^+$($S^-$)
 and $\in H(\overline{S^+})$ ($\in H(\overline{S^-}$).
 We know that \cite{lu, mus}
\begin{equation}
\left\{
\begin{array}{ll}
\!\!\!\!\!\!\!\!&\mathbf{C}^{\pm}[\varphi]\in A^{\pm}H,\\[3mm]
\!\!\!\!\!\!\!\!&\mathbf
{C}[\varphi](\infty)=\lim\limits_{z\rightarrow\infty}\mathbf
{C}[\varphi](z)=0.
\end{array}
\right.
\end{equation}

A function $F(z)$ is said to be sectionally holomorphic with $L$ as
its jump curve, if it is holomorphic in $S^+$ and $S^-$, probably
possesses a pole at infinity and
 has the finite boundary values $F^+(t)$ and $F^-(t)$
  when $z$ tends to any point $t\in L$ from $S^+$
 and $S^-$ respectively.
 Sometimes it is of convenience that we introduce two
projection functions
\begin{equation}
 {F}^{\pm}(z)=\left\{
\begin{array}{ll}
{F}(z),\hspace{4mm}&z\in S^{\pm},\\[3mm]
F^{\pm}(t),\hspace{4mm}&z=t\in L.
\end{array}
\right.
\end{equation}
Obviously, $\mathbf{C}[\varphi]|_{S^+\cup S^-}$ is sectionally
holomorphic with $L$ as its jump curve.

The Riemann boundary value problem, or simply RP,
  is to find a sectionally holomorphic function
 $\Phi$ with $L$ as
 its jump curve satisfying the Riemann boundary value condition
 \begin{equation}
 \Phi^+(t)=G(t)\Phi^-(t)+g(t),\hspace{5mm}t\in L,
 \end{equation}
 where $G$ and $g$ are given functions on $L$, both $\in H(L)$, $G(t)\not=0$ for
 $t\in L$ (normal type). In RP (2.39), $G$ and $g$ are called  the discontinuous factor and the
 jump function, or simply factor and jump respectively. If $\Phi$ is required of order at
 most $m$ at infinity,
 then the problem is denoted by $R_m$.
The complete solution of RP can be found in \cite{gak,lu,mus}. The
RPs treated in the present paper are all the simplest, which can be
directly solved by (2.36).

\begin{exa}{\rm Let $\partial \mathbb{D}$ be the unit circle
oriented counter--clockwise and weight function $w\in H(\partial
\mathbb{D})$ be a positive valued function. Find a sectionally
holomorphic function $D$ with $\partial \mathbb{D}$ as its jump
curve such that $D(z)\not=0$ for $z\in S^+\cup S^-$ and satisfying
the following condition
\begin{equation}
 \left\{
\begin{array}{ll}
&\hspace{-4mm}D^+(t)D^-(t)=w(t),\hspace{6mm}t\in \partial \mathbb{D},\\[3mm]
&\hspace{-4mm}D(\infty)=1.
\end{array}\right.
\end{equation}
Let
\begin{equation} X(z)=\exp\left\{\mathbf{C}[\log
w](z)\right\},\hspace{6mm}z\not\in \partial \mathbb{D}.
\end{equation} Obviously, by
(2.37) we know that
\begin{equation} D(z)=\left\{
\begin{array}{ll}
X^+(z)=\exp\left\{\mathbf{C}[\log w](z)\right\},\hspace{6mm}&z\in
S^+=\{z:\,|z|<1\},
\\[5.4mm]
(X^-)^{-1}(z)=\exp\left\{-\mathbf{C}[\log w](z)\right\},\hspace{6mm}
&z\in S^-=\{z:\,|z|>1\}
\end{array}
\right.
\end{equation} is a solution of (2.40). In fact, it is the
unique solution of (2.40), because (2.40) has only the solutions
$D(z)\equiv1$ while $w\equiv1$ by the Liouville's theorem. By the
way, in the general theory of Riemann boundary value problems (2.40)
is just a homogeneous $R_0$ problem, i.e., $g\equiv0$ in (2.39),
with $\Phi^-=(D^-)^{-1}$ and $\Phi^+=D^+$ under the  condition
$D(\infty)=1$.
 Since the index $\kappa=0$ in this case, $\Phi$ is certainly
 the canonical function $X$ of (2.39) (for example, see \cite{lu}).
}
\end{exa}

The other kind of Riemann boundary value problems, so--called
Riemann boundary value  problems for matrix valued functions, or
simply MRPs,
 will also be used in the sequel. To do so, we introduce
some terminologies at first.

Let
\begin{equation}
 \mathbf\Phi(z)=\left(
\begin{array}{rr}
\Phi_{1,1}(z)&\Phi_{1,2}(z)\\[2mm]
\Phi_{2,1}(z)&\Phi_{2,2}(z)
\end{array}
\right)
\end{equation} be a $(2\times2)$ matrix valued function
defined on a set $\Omega$ in the complex plane, where each element
$\Phi_{j, k}$ is a function defined on $\Omega$. Whenever a property
such as continuity, analyticity, etc. is ascribed of $\mathbf\Phi$,
it is clear that in fact all the element functions $\Phi_{j, k}$
possess the cited property. So the meanings of $\mathbf\Phi\in
H(L)$, $\mathbf\Phi$ is sectionally holomorphic with $L$ as its jump
curve, etc. are obvious. In particular,

\begin{equation} \lim\limits_{z\rightarrow\infty}\mathbf\Phi(z)=\mathbf
a=:\left(
\begin{array}{rr}
a_{1,1}&a_{1,2}\\[2mm]
a_{2,1}&a_{2,2}
\end{array}
\right)
\end{equation} means
\begin{equation}
\lim\limits_{z\rightarrow\infty}\Phi_{j,k}(z)=a_{j,k},\hspace{5mm}j,k=1,2,
\end{equation} where $\mathbf a$ is a complex valued matrix. But we would
rather treat (2.44) as the convergence  in the sense of the norm

\begin{equation}
\|\mathbf\Phi(z)\|=\sum\limits_{j,k=1}^{2}\left|\Phi_{j,k}(z)\right|.
\end{equation} Sometimes we also use the following norm

\begin{equation}
\|\mathbf\Phi\|_{\Omega}=\sum\limits_{j,k=1}^{2}\left\|\Phi_{j,k}\right\|_{\Omega}
\hspace{4mm}\mathrm{with}\hspace{4mm}
\left\|\Phi_{j,k}\right\|_{\Omega}=
\sup\Bigl\{|\Phi_{j,k}(z)|,\,\,z\in\Omega\Bigr\}.
\end{equation}

Now we state Riemann boundary value problems
 for
matrix valued functions as follows: Find a sectionally holomorphic
matrix valued function $\mathbf\Phi$, with $L$ as
 its jump curve satisfying the Riemann boundary value condition
\begin{equation}
 \mathbf\Phi^+(t)=\mathbf G(t)\mathbf\Phi^-(t)+\mathbf g(t),\hspace{5mm}t\in L,
 \end{equation}
 where $\mathbf G$ and $\mathbf g$ are given $(2\times2)$ matrix valued
 functions on $L$, both $\in H(L)$,
  $\det\mathbf G(t)\not=0$ for
 $t\in L$ (normal type). $G$ and $g$ are respectively called the
 discontinuous factor and jump matrices. In addition,
 we say that
 $\mathbf\Phi$ is of order $m$ at infinity, denoted as
 $
\mathrm{Ord}(\mathbf\Phi,\infty)=m$, if $m=\max\{m_{j,k}:\,
j,k=1,2\}$
 where $m_{j,k}=\mathrm{Ord}\left(\Phi_{j,k},\infty\right)$ is the order of $\Phi_{j,k}$ at
 infinity. If it is required that
 \begin{equation}
\mathrm{Ord}(\mathbf\Phi,\infty)\leq m,
\end{equation}
 then the problem is denoted by $MR_m$.

 The complete theory of MRPs can be established since they can be transformed into
 Riemann boundary value problems for the system of analytic
functions (simply, SRPs) whereas the complete theory of the latter
can be found in \cite{gak, lu, mus}. The transformation process is
in detail stated in \cite{dd}.

As the rule  mentioned above, the principal part of the sectionally
holomorphic matrix valued function $\mathbf\Phi$ given in (2.43) at
infinity is

\begin{equation} \mathbf{P.P}[\mathbf\Phi,\infty](z)=\left(
\begin{array}{rr}
\mathbf{P.P}[\Phi_{1,1},\infty](z)&\mathbf{P.P}[\Phi_{1,2},\infty](z)\\[2mm]
\mathbf{P.P}[\Phi_{2,1},\infty](z)&\mathbf{P.P}[\Phi_{2,2},\infty](z)
\end{array}
\right),
\end{equation} where $\mathbf{P.P}[\Phi_{j,k},\infty]$ are
the principal parts of the sectionally holomorphic
 functions $\Phi_{j,k}$ at infinity. While we change the requirement
 (2.49) to the condition for the behavior of the principal part at infinity
 \begin{equation}
\mathbf{P.P}[\mathbf\Xi\mathbf\Phi,\infty](z)=\mathbf{\Pi}(z)=\left(
\begin{array}{cc}
\Pi_{1,1}(z)&\Pi_{1,2}(z)\\[3mm]
\Pi_{2,1}(z)&\Pi_{2,2}(z)
\end{array}
\right)
\end{equation} where $\mathbf\Xi$ and $\mathbf{\Pi}$ are
respectively a given analytic matrix valued function on some
neighborhood of the point at infinity and a polynomial matrix valued
function, the MRP (2.48) gives rise to serious difficulty. The
general discussion for it will be postponed to another paper in
order to avoid departing from the subject. In the present paper, we
only deal with some simple homogeneous MRPs which will be used in
the sequel.

\begin{exa}
{\rm We consider the simplest MRP, which is a skew type MRP for a
sectionally holomorphic matrix valued function $\mathbf M$ with
$\partial \mathbb{D}$ as
 its jump curve:
\begin{equation}
\left\{
\begin{array}{l}
\mathbf M^+(t)=\left(\begin{array}{cc}
0&-1\\[2mm]
1&0
\end{array}
\right)\mathbf M^{-}(t), \hspace{3mm}t\in \partial \mathbb{D}
\\[8mm]
\mathbf{P.P}[\mathbf M,\infty](z)=I,
\end{array}
\right.
\end{equation}
 where $I$ is the $(2\times2)$ identity matrix. By a
direct calculation and using Liouville's theorem, we easily know
that

\begin{equation}
 \mathbf{M}(z)= \left\{
\begin{array}{cl}
\left(\begin{array}{cc}0&-1\\[2mm]1&0\end{array}\right),
& \hspace{4mm}|z|<1,\\[10mm]
\left(\begin{array}{cc}
1&0\\[2mm]
0&1
\end{array}
\right), &\hspace{4mm} |z|>1
\end{array}
\right. \end{equation} is the unique solution of MRP (2.52). }
\end{exa}

\begin{exa}{\rm
Let $L$ be a set of smooth closed curves $L_1,L_2,\ldots,L_n$ in the
complex plane, non-intersecting to each other and positively
oriented for each $L_j$. We consider the following MRP:
\begin{equation}
\left\{
\begin{array}{ll}
\mathbf{R}^{+}(t)=\mathbf{V}(t)\mathbf{R}^{-}(t), \hspace{4mm}t\in L,\\[3mm]
\mathbf{P.P}[\mathbf{R},\infty](z)=I,
\end{array}
\right.
\end{equation}
 where $\mathbf{V}\in H(L)$ and
$\det{\mathbf{V}}(t)\not=0$ for any $t\in L$. Moreover assume
$\mathbf{V}$ is analytic on $L$, i.e., analytic on some open
neighborhood $\Omega$ of $L$. Under these assumptions, we can show
that the solution $\mathbf{R}$ of (2.54) is controlled  by the
factor matrix $\mathbf{V}$ in certain sense. }
\end{exa}
 \begin{lem}[Du and Du \cite{dd}] There
exist constants $C>0$ and $\delta>0$ such that
\begin{equation}
\left\|\mathbf{R}-I\right\|_{\mathcal{C}\setminus L}\leq
C\left\|\mathbf{V}-I\right\|_{\Omega}
\hspace{4mm}while\hspace{4mm}\left\|\mathbf{V}-I\right\|_{\Omega}<\delta.
\end{equation}
\end{lem}
\begin{rem}{\rm When $L$ is a simple closed contour, this result is due to
Aptekarev \cite{apt} (also see \cite{kui}).}\end{rem}

\section{Riemann-Hilbert Analysis for Orthogonal Trigonometric Polynomials}

In the present section, we only consider asymptotics for the first
orthonormal Laurent polynomials $\mu_{n}$ on the unit circle in
$\mathbb{C}\setminus\{0\}$ by the Riemann-Hilbert approach. As a
consequence, we easily obtain asymptotics for the first orthonormal
trigonometric polynomials $\pi_{n}(\theta)$ and $\sigma_{n}(\theta)$
on $[0,\,2\pi)$. Similarly, asymptotics for the second orthonormal
Laurent or trigonometric polynomials on the unit circle can be also
obtained by the Riemann-Hilbert approach or in terms of the basis
transform formula between the first orthonormal Laurent or
trigonometric polynomials on the unit circle and the second ones.

\subsection{Characterization of Orthogonal Trigonometric Polynomials}

We begin with the characterization of the first orthogonal Laurent
or trigonometric polynomials on the unit circle.

The characterization is the following MRP
\begin{equation}
\left\{\begin{array}{l} \mathbf{Y}^{+}(t)=\left(
\begin{array}{cc}
      1 & 0 \\
    t^{-2n}w(t)& 1
    \end{array}\right)\mathbf{Y}^{-}(t),\hspace{4mm}t\in \partial \mathbb{D},\\[6mm]
\mathbf{P.P}[\mathbf\Xi\mathbf{Y},\infty](z)=I,\\[6mm]
Y_{1,1}(0)=Y_{1,2}(0)=0
\end{array}
\right.
\end{equation} with
\begin{equation}
\mathbf\Xi(z)=\left(\begin{array}{cc}
z^{-2n}&0\\[2mm]
0&z^{2n-1}
\end{array}
\right),
\end{equation}
where the unit circle $\partial \mathbb{D}$ is counter-clockwise
oriented and $w\in H(\partial \mathbb{D})$, which is a positive
weight function on the unit circle.

\begin{thm}
The MRP (3.1) is uniquely solvable and its unique solution is
\begin{equation}
\begin{pmatrix}
       z^{n}[\lambda_{1,n}a_{n}\sigma_{n}(z)+\lambda_{2,n}b_{n}\pi_{n}(z)] &
       z^{n}[\lambda_{3,n-1}a_{n-1}\sigma_{n-1}(z)+\lambda_{4,n-1}b_{n-1}\pi_{n-1}(z)]
       \\[2mm]
       \mathbf{C}\big[\tau^{n}\big(\lambda_{1,n}a_{n}\sigma_{n}+\lambda_{2,n}b_{n}\pi_{n}\big)\big](z) &
       \mathbf{C}\big[\tau^{n}\big(\lambda_{3,n-1}a_{n-1}\sigma_{n-1}+\lambda_{4,n-1}b_{n-1}\pi_{n-1}\big)\big](z)\\
     \end{pmatrix},
\end{equation}
where $\sigma_{n}(z),\pi_{n}(z)$ are respectively the first
orthonormal Laurent polynomials of order $2n$ and $2n-1$ on the unit
circle, $a_{n}, b_{n}, \beta_{n}$ are given in (2.17),
$\lambda_{1,n}=1$, $\lambda_{2,n}=\beta_{n}+i$,
$\lambda_{3,n-1}=-\frac{1}{2}a^{-2}_{n-1}(1+\beta_{n-1}i)$,
$\lambda_{4,n-1}=\frac{1}{2}b^{-2}_{n-1}i$, and the Cauchy singular
integral operator $\mathbf{C}$ is given by (2.35).
\end{thm}

\begin{proof}
The boundary value conditions in (3.1) can be explicitly written as

\begin{align}
Y_{1,1}^+(t)=&\,\,Y_{1,1}^{-}(t), \hspace{6mm}t\in \partial \mathbb{D},\\
Y_{2,1}^+(t)=&\,\,t^{-2n}w(t)Y_{1,1}^{-}(t)+Y_{2,1}^{-}(t), \hspace{6mm}t\in \partial \mathbb{D},\\
Y_{1,2}^+(t)=&\,\,Y_{1,2}^{-}(t), \hspace{6mm}t\in \partial \mathbb{D},\\
Y_{2,2}^+(t)=&\,\,t^{-2n}w(t)Y_{1,2}^{-}(t)+Y_{2,2}^{-}(t),
\hspace{6mm}t\in \partial \mathbb{D}.
\end{align}
Clearly write the condition for the behavior of the principal part
at infinity by
\begin{equation}
\left(\begin{array}{cc}
\mathbf{P.P}[z^{-2n}Y_{1,1},\infty](z)&\mathbf{P.P}[z^{-2n}Y_{1,2},\infty](z)\\[3mm]
\mathbf{P.P}[z^{2n-1}Y_{2,1},\infty](z)&\mathbf{P.P}[z^{2n-1}Y_{2,2},\infty](z)
\end{array}
\right)=\left(\begin{array}{cc}
1&0\\[3mm]
0&1
\end{array}
\right).
\end{equation}
Obviously, $Y_{1,1}(z)$ is a monic polynomial of order $2n$. Note
that $Y_{1,1}(0)=0$, set
\begin{equation}
Y_{1,1}(z)=z^{n}P_{1,1}(z), \,\,\,\,\,P_{1,1}\in H^{L}_{2n},
\end{equation}
where $H^{L}_{2n}$ denotes the span of the first $2n$ elements of
the ordered set $\{1, z, z^{-1}, \ldots, z^{n}$, $z^{-n}, \ldots\}$,
i.e., $P_{1,1}$ is a Laurent polynomial spaned by $\{1, z, z^{-1},
\ldots, z^{n-1}, z^{-n+1}, z^{n}\}$ with 1 as the coefficient of
$z^{n}$.

Substituting (3.9) into (3.5), we have
\begin{align}
Y_{2,1}(z)&=\frac{1}{2\pi i}\int_{\partial
\mathbb{D}}\frac{\tau^{-n}P_{1,1}(\tau)w(\tau)}{\tau-z}d\tau\\
&=\frac{1}{2\pi i}\int_{\partial
\mathbb{D}}\frac{\tau^{-n+1}P_{1,1}(\tau)w(\tau)}{\tau-z}\frac{d\tau}{\tau}.
\end{align}
Note that for sufficiently large $z$,
\begin{equation}
\frac{1}{\tau-z}=-\sum_{k=0}^{\infty}\frac{\tau^{k}}{z^{k+1}}.
\end{equation}
So by (3.8), (3.11) and (3.12), it is easy to see that
\begin{equation}
\frac{1}{2\pi i}\int_{\partial
\mathbb{D}}\tau^{-n+1+k}P_{1,1}(\tau)w(\tau)\frac{d\tau}{\tau}=0,\,\,k=0,1,\ldots,2n-2.
\end{equation}
It shows that
\begin{equation}
\langle\tau^{\pm j},
P_{1,1}\rangle_{\mathbb{R}}=0,\,\,\,\,j=0,1,\ldots,n-1.
\end{equation}
Therefore,
\begin{equation}
P_{1,1}(z)=\lambda_{1,n}a_{n}\sigma_{n}(z)+\lambda_{2,n}b_{n}\pi_{n}(z),
\end{equation}
where $\lambda_{1,n},\lambda_{2,n}$ are some complex constants only
depending on $n$, $\sigma_{n}(z),\pi_{n}(z)$ are respectively the
first orthogonal Laurent polynomials of order $2n$ and $2n-1$ on the
unit circle, and $a_{n}, b_{n}$ are given in (2.17). Since the
coefficient of $z^{n}$ in $P_{1,1}(z)$ is 1 and $P_{1,1}\in
H^{L}_{2n}$ defined as above, we have

\begin{equation}
\begin{cases}
\big(\frac{1}{2}-\frac{\beta_{n}}{2i}\big)\lambda_{1,n}+\frac{1}{2i}\lambda_{2,n}=1,\\[2mm]
\big(\frac{1}{2}+\frac{\beta_{n}}{2i}\big)\lambda_{1,n}-\frac{1}{2i}\lambda_{2,n}=0,
\end{cases}
\end{equation}
where $\beta_{n}$ is given in (2.17). Then
\begin{equation}
\lambda_{1,n}=1,\,\,\,\,\lambda_{2,n}=\beta_{n}+i.
\end{equation}
Hence,
\begin{equation}
Y_{1,1}(z)=z^{n}[a_{n}\sigma_{n}(z)+(\beta_{n}+i)b_{n}\pi_{n}(z)]
\end{equation}
and
\begin{equation}
Y_{2,1}(z)=\frac{1}{2\pi i}\int_{\partial
\mathbb{D}}\frac{\tau^{n}[a_{n}\sigma_{n}(\tau)+(\beta_{n}+i)b_{n}\pi_{n}(\tau)]w(\tau)}{\tau-z}d\tau.
\end{equation}

Analogously, $Y_{1,2}(z)$ is a polynomial of order at most $2n-1$
which follows from (3.6) and (3.8). All and the same, note that
$Y_{1,2}(0)=0$, let
\begin{equation}
Y_{1,2}(z)=z^{n}P_{1,2}(z),\,\,\,\,P_{1,2}\in H_{2n-1}^{L},
\end{equation}
i.e., where $P_{1,2}$ is a Laurent polynomial spaned by $\{1, z,
z^{-1}, \ldots, z^{n-1}, z^{-n+1}\}$.

Put (3.20) into (3.7), we get
\begin{align}
Y_{2,2}(z)&=\frac{1}{2\pi i}\int_{\partial
\mathbb{D}}\frac{\tau^{-n}P_{1,2}(\tau)w(\tau)}{\tau-z}d\tau\\
&=\frac{1}{2\pi i}\int_{\partial
\mathbb{D}}\frac{\tau^{-n+1}P_{1,2}(\tau)w(\tau)}{\tau-z}\frac{d\tau}{\tau}.
\end{align}

By (3.8), (3.12) and (3.22),
\begin{equation}
\frac{1}{2\pi i}\int_{\partial
\mathbb{D}}\tau^{-n+1+k}P_{1,2}(\tau)w(\tau)\frac{d\tau}{\tau}=0,\,\,k=0,1,\ldots,2n-3
\end{equation}
and
\begin{equation}
\frac{1}{2\pi i}\int_{\partial
\mathbb{D}}\tau^{n-1}P_{1,2}(\tau)w(\tau)\frac{d\tau}{\tau}=-1.
\end{equation}

(3.23) and (3.24) are equivalent to
\begin{equation}
\langle\tau^{\pm j},
P_{1,2}\rangle_{\mathbb{R}}=0,\,\,\,\,j=0,1,\ldots,n-2
\end{equation}
and
\begin{equation}
\begin{cases}
\langle\tau^{n-1},\, P_{1,2}\rangle_{\mathbb{R}}=-1,\\[1mm]
\langle\tau^{-n+1},\, P_{1,2}\rangle_{\mathbb{R}}=0.\\
\end{cases}
\end{equation}

From (3.25),
\begin{equation}
P_{1,2}(z)=\lambda_{3,n-1}a_{n-1}\sigma_{n-1}(z)+\lambda_{4,n-1}b_{n-1}\pi_{n-1}(z),
\end{equation}
where $\lambda_{3,n-1},\lambda_{4,n-1}$ are some complex constants
only depending on $n-1$, $\sigma_{n-1}(z)$, $\pi_{n-1}(z)$ are
respectively the first orthogonal Laurent polynomials of order
$2n-2$ and $2n-3$ on the unit circle, and $a_{n-1}, b_{n-1}$ are
given in (2.17).

Applying (3.27) to (3.26), we obtain
\begin{equation}
\begin{cases}
\langle\mu^{(0)}_{2n-2},\,
a_{n-1}\sigma_{n-1}\rangle_{\mathbb{R}}\lambda_{3,n-1}
+\langle\mu^{(0)}_{2n-2},\, b_{n-1}\pi_{n-1}\rangle_{\mathbb{R}}\lambda_{4,n-1}=-1/2,\\[1mm]
\langle\mu^{(0)}_{2n-3},\,
a_{n-1}\sigma_{n-1}\rangle_{\mathbb{R}}\lambda_{3,n-1}
+\langle\mu^{(0)}_{2n-3},\, b_{n-1}\pi_{n-1}\rangle_{\mathbb{R}}\lambda_{4,n-1}=-1/2i.\\
\end{cases}
\end{equation}
Equivalently,
\begin{equation}
\begin{cases}
a_{n-1}^{2}\lambda_{3,n-1}
+\beta_{n-1}b_{n-1}^{2}\lambda_{4,n-1}=-1/2,\\[1mm]
\,\,\,\,\,\hspace{26mm}b_{n-1}^{2}\lambda_{4,n-1}=-1/2i.
\end{cases}
\end{equation}
Therefore,
\begin{equation}
\lambda_{3,n-1}=-\frac{1}{2}a_{n-1}^{-2}(1+\beta_{n-1}i),
\,\,\,\,\lambda_{4,n-1}=\frac{1}{2}b_{n-1}^{-2}i.
\end{equation}
So
\begin{equation}
Y_{1,2}(z)=-\frac{1}{2}z^{n}[a^{-1}_{n-1}(1+\beta_{n-1}i)\sigma_{n-1}(z)-ib^{-1}_{n-1}\pi_{n-1}(z)]
\end{equation}
and
\begin{equation}
Y_{2,2}(z)=-\frac{1}{4\pi i}\int_{\partial
\mathbb{D}}\frac{\tau^{n}[a^{-1}_{n-1}(1+\beta_{n-1}i)\sigma_{n-1}(\tau)-ib^{-1}_{n-1}\pi_{n-1}(\tau)]w(\tau)}{\tau-z}d\tau.
\end{equation}

By the uniqueness of $\sigma_{n}$ and $\pi_{n}$, we complete the
proof of Theorem 3.1.
\end{proof}

\begin{rem}
In the above proof, we get some complex constants $\lambda_{1,n}$,
$\lambda_{2,n}$, $\lambda_{3,n}$ and $\lambda_{4,n}$ which only
depend on $n$. In the sequel, in order to independently obtain
asymptotics of $\sigma_{n}$ and $\pi_{n}$, we must consider the
following determinant

\begin{equation}
\Lambda_{n}=\begin{vmatrix}
              \lambda_{1,n} & \lambda_{2,n} \\
              -\lambda_{3,n} & -\lambda_{4,n} \\
            \end{vmatrix}.
\end{equation}
By straightforward calculations,
\begin{equation}
\Lambda_{n}=-\frac{1}{2}[a_{n}^{-2}(1+\beta_{n}^{2})+b_{n}^{-2}]i\neq0.
\end{equation}
In what follows, one will find that this determinant is extremely
important, useful and powerful. In section 5, we will give another
explanation on $\Lambda_{n}$ in terms of OPUC.
\end{rem}

\subsection{Steepest Descent Analysis for Characterization}

In the last section, we have used a MRP to characterize the first
orthogonal Laurent polynomials on the unit circle with positive and
H\"older continuous weight functions. Obviously, they become the
first orthogonal trigonometric polynomials by restricting them to
the unit circle. In this section, in order to obtain their
asymptotics, we analyze the characterization by steepest descent
method originally introduced by Deift and Zhou in \cite{dz1} which
usually consists of a series of explicit transforms. In the case of
the present paper, the strategy of the steepest descent method is
extremely similar to the one used in \cite{dd} with some
modifications. It depends on a series of transforms as follows
\begin{equation}
\mathbf{Y}\mapsto\mathbf{D}\mapsto\mathbf{F}\mapsto\mathbf{S}\mapsto\mathbf{R},
\end{equation}
where $\mathbf{Y}$ is given by the characterization (3.3) in the
last section and $\mathbf{R}$ is given in Example 2.3 in Section 2
which will be given in greater detail in this section. Then the
asymptotics of $\mathbf{Y}$ is done by  applying Lemma 2.4 to
$\mathbf{R}$. Each of these  transforms brings us closer to our
claim. From now on, we always suppose that the weight function $w$
is positive and analytic on the unit circle. That is, $w$ is
analytic in a open neighborhood of $\partial \mathbb{D}$ and
positive on $\partial \mathbb{D}$.

{\bf First transform} $\mathbf{Y}\mapsto\mathbf{D}$. The aim of this
transform is very clear, which  is  to make that the new MRP is
normalized at infinity. Let
 \begin{equation}
\mathbf{D}(z)=\left\{\begin{array}{ll} \left(\begin{array}{cc}
  D(z)& 0 \\[2mm]
  0 & D^{-1}(z)
\end{array}
\right) \mathbf{Y}(z),\hspace{2mm} & |z|<1, \\[14mm]
\left(\begin{array}{cc}
  z^{-2n}D(z)& 0 \\[2mm]
  0 & z^{2n-1}D^{-1}(z)
\end{array}
\right) \mathbf{Y}(z),\hspace{2mm} & |z|>1,
  \end{array}
 \right.
  \end{equation}
where $D$ and $\mathbf{Y}$ are respectively given in Example 2.1 and
by the characterization in the last section. Then, we
straightforward get from (3.1), that $\mathbf{D}$ is a sectionally
holomorphic matrix valued function with $\partial \mathbb{D}$ as its
jump curve, satisfying the following MRP

\begin{equation}
   \left\{
   \begin{array}{ll}
   \mathbf{D}^{+}(t)=\left(
   \begin{array}{cc}
         \displaystyle\frac{t^{2n}(D^+)^{2}(t)}{w(t)} &0 \\[3mm]
      1& \displaystyle\frac{t^{-2n+1}(D^-)^{2}(t)}{w(t)}
    \end{array} \right)\mathbf{D}^-(t),\hspace{4mm}t\in\partial \mathbb{D},\\[12mm]
    \mathbf{P.P}[\mathbf{D},\infty](z)=I,\\[5mm]
    D_{1,1}(0)=D_{1,2}(0)=0.

\end{array}
\right.
\end{equation}

\begin{rem}{\rm The MRP (3.37) is uniquely solvable because it is
equivalent to the MRP (3.1) under the transform (3.36).}
\end{rem}

{\bf Second transform} $\mathbf{D}\mapsto\mathbf{F}.$ This transform
is an additional one by comparing the steepest descent method used
in \cite{dd}. Note that the determinant of the factor matrix for
$\mathbf{D}$ is $t$ but not identically equal to 1, namely,
\begin{equation}
\begin{vmatrix}
  \displaystyle\frac{t^{2n}(D^+)^{2}(t)}{w(t)} & 0 \\
  1 & \displaystyle\frac{t^{-2n+1}(D^-)^{2}(t)}{w(t)} \\
\end{vmatrix}=t\neq 1,\,\,\,\,\,\mathrm{as}\,\, 1\neq t\in \partial \mathbb{D}.
\end{equation}
However, the determinants of the factor matrices in the steepest
descent method are usually equal to 1. This is conveniently
necessary for the design on the steepest descent method. Such aim
will be attained by the discussing transform. To do so, let

\begin{equation}
\mathbf{F}(z)=\begin{cases}
\begin{pmatrix}
  \displaystyle\frac{D_{1,1}(z)}{z} & \displaystyle\frac{D_{1,2}(z)}{z} \\[4mm]
  D_{2,1}(z) & D_{2,2}(z) \\
\end{pmatrix},\hspace{4mm} 0\neq|z|<1,\\[10mm]
\begin{pmatrix}
  \lim\limits_{z\rightarrow 0}\displaystyle\frac{D_{1,1}(z)}{z} & \lim\limits_{z\rightarrow 0}
  \displaystyle\frac{D_{1,2}(z)}{z} \\[4mm]
  D_{2,1}(0) & D_{2,2}(0) \\
\end{pmatrix},\hspace{4mm} z=0,\\[10mm]
\mathbf{D}(z),\hspace{4mm} |z|>1,
\end{cases}
\end{equation}
then from (3.37), $\mathbf{F}$ is a sectionally holomorphic matrix
valued function with $\partial \mathbb{D}$ as its jump curve,
satisfying the following MRP

\begin{equation}
   \left\{
   \begin{array}{ll}
   \mathbf{F}^{+}(t)=\left(
   \begin{array}{cc}
         \displaystyle\frac{t^{2n-1}(D^+)^{2}(t)}{w(t)} &0 \\[3mm]
      1& \displaystyle\frac{t^{-2n+1}(D^-)^{2}(t)}{w(t)}
    \end{array} \right)\mathbf{F}^-(t),\hspace{4mm}t\in\partial \mathbb{D},\\[12mm]
    \mathbf{P.P}[\mathbf{F},\infty](z)=I.
\end{array}
\right.
\end{equation}
It must be noted that MRP (3.40) is just the first transformed MRP
\cite{dd} with $2n-1$ here in place of $n$ there. So we can directly
apply analogous steepest descent transforms used in \cite {dd} to
construct the sequent ones in the present paper, i.e., the steepest
descent transforms $\mathbf{F}\mapsto\mathbf{S}\mapsto\mathbf{R}$.

\begin{rem}{\rm The MRP (3.40) is uniquely solvable because it is
equivalent to the MRP (3.37) under the transform (3.39).}
\end{rem}

Just as the above statement, the detail policy of the following
transforms can be found in \cite{dd}.

{\bf Third transform} $\mathbf{F}\mapsto\mathbf{S}.$ This transform
is that the jump curve $\partial \mathbb{D}$ of $\mathbf{F}$ will be
artificially enlarged so that the discontinuous factor matrix on the
appended part of the new jump curve can fulfill the condition in the
assertion (2.55). It is based on a convenient factorization of the
matrix
\begin{align}
&\begin{pmatrix}
  \displaystyle\frac{t^{2n-1}(D^{+})^2(t)}{w(t)} & 0 \\
  1 & \displaystyle\frac{t^{-2n+1}(D^{-})^2(t)}{w(t)} \\
\end{pmatrix}\\
=&\begin{pmatrix}
  1 & \displaystyle\frac{t^{2n-1}(D^{+})^{2}(t)}{w(t)} \\
  0 & 1 \\
\end{pmatrix}
\begin{pmatrix}
  0 & -1 \\
  1& 0 \\
\end{pmatrix}
\begin{pmatrix}
  1 & \displaystyle \frac{t^{-2n+1}(D^{-})^{2}(t)}{w(t)} \\
  0 & 1 \\
\end{pmatrix},\,\,\,\,t\in \partial \mathbb{D}.\nonumber
\end{align}

Let \begin{equation}
\mathbf{S}(z)= \left\{\begin{array}{ll}
\mathbf{F}(z),\hspace{3mm}&z\in S^+_{\sharp2}\cup S^{-}_{\sharp1},\\[4mm]
\left(\begin{array}{cc}
    1 &\displaystyle\frac{z^{-2n+1}D^2(z)}{w(z)} \\[3mm]
    0 & 1
  \end{array}
  \right)\mathbf{F}(z),\hspace{3mm}&z\in S^{-}_{\sharp2},\\[10mm]
\left(\begin{array}{cc}
    1 &-\displaystyle\frac{z^{2n-1}D^2(z)}{w(z)} \\[3mm]
    0 & 1
  \end{array}
  \right)\mathbf{F}(z),\hspace{3mm}&z\in S^{+}_{\sharp1}
\end{array}
\right.
\end{equation}
and
 \begin{equation}
\mathbf{\Upsilon}(t)=\left\{
\begin{array}{ll}
\left(\begin{array}{cc}
    1 &\displaystyle\frac{t^{-2n+1}(D^-)^{2}(t)}{w(t)} \\[3mm]
    0 & 1
  \end{array}
  \right),\hspace{3mm}&t\in \partial \mathbb{D}_{1+\epsilon},\\[10mm]
   \left(\begin{array}{cc}
    0 & 1 \\[3mm]
    -1 & 0
  \end{array}\right),\hspace{3mm}&t\in \partial \mathbb{D},\\[12mm]
   \left(
    \begin{array}{cc}
  1 & \displaystyle\frac{t^{2n-1}(D^{+})^{2}(t)}{w(t)}\\[3mm]
 0 & 1
  \end{array}
  \right),\hspace{3mm}&t\in\partial \mathbb{D}_{1-\epsilon},
  \end{array}
  \right.
\end{equation}
$\epsilon$ be a positive real constant and sufficiently small,
\begin{equation} \partial \mathbb{D}_{1-\epsilon}=\{z:\,
|z|=1-\epsilon\},\hspace{5mm}\partial \mathbb{D}_{1+\epsilon}=\{z:\,
|z|=1+\epsilon\},
\end{equation}
\begin{equation} S^{+}_{\sharp2}=\{z:\,|z|>1+\epsilon,
z=\infty\},\hspace{5mm} S^{+}_{\sharp1}=\{z:\, 1-\epsilon<|z|<1\},
\end{equation}
and
\begin{equation} S^{-}_{\sharp2}=\{z:\,
1<|z|<1+\epsilon\},\hspace{5mm}S^{-}_{\sharp1}=\{z:\,
|z|<1-\epsilon\}.
\end{equation}
In the above, $\partial \mathbb{D}_{1-\epsilon}$ and $\partial
\mathbb{D}_{1+\epsilon}$ are clockwise oriented while the unit
circle $\partial \mathbb{D}$ is counter--clockwise oriented,
$L_\sharp=\partial \mathbb{D}_{1-\epsilon}+\partial
\mathbb{D}+\partial \mathbb{D}_{1+\epsilon}$ divides the extended
complex plane into two parts $S^{+}_{\sharp}=S^{+}_{\sharp1}\cup
S^{+}_{\sharp2}$ and $S^{-}_{\sharp}=S^{-}_{\sharp1}\cup
S^{-}_{\sharp2}$. Here, we have assumed that the weight function $w$
is analytic in
\begin{equation}
U(\epsilon)=\{z:\,1-3\epsilon<|z|<1+3\epsilon\}
\,\,\,\,\mathrm{with}\,\,\,\,0<\epsilon<1/3.
\end{equation}

Then, it is easy to directly verify that $\mathbf{S}$ is a solution
of the following MRP:
\begin{equation} \left\{
\begin{array}{ll}
\mathbf{S}^-(t)=\mathbf{\Upsilon}(t)\mathbf{S}^+(t),\hspace{3mm}t\in L_\sharp,\\[3mm]
\mathbf{P.P}[\mathbf{S},\infty](z)=I.
\end{array}
\right.
\end{equation}

\begin{rem}{\rm The MRP (3.48) is uniquely solvable because it is
equivalent to the MRP (3.40) under the transform (3.42).}
\end{rem}

{\bf Fourth transform} $\mathbf{S}\mapsto\mathbf{R}$.
 The effect of this transform is to take  out the discontinuity on $\partial \mathbb{D}$ such that
$\mathbf{R}$ may be analytic across the unit circle $\partial
\mathbb{D}$. Thus we get a MRP fulfilling the conditions in the MRP
(2.54) and assertion (2.55). To do so, set
 \begin{equation}
\mathbf{R}(z)=\mathbf{M}^{-1}(z)\mathbf{S}(z),\hspace{4mm}z\in
\mathcal{C}\setminus L_{\sharp}.
\end{equation} In fact, by using
Example 2.2 we have that
\begin{equation}
\left(\mathbf{M}^+\right)^{-1}\!\!(t)\,\mathbf{S}^+(t)=\left(\mathbf{M}^-\right)^{-1}\!\!(t)
\,\mathbf{S}^-(t),\hspace{6mm}t\in\partial \mathbb{D}.
\end{equation} Thus, we
get the sectionally holomorphic matrix valued function $\mathbf{R}$
with the jump curve $L=(-\partial \mathbb{D}_{1+\epsilon})+\partial
\mathbb{D}_{1-\epsilon}$ where
 $-\partial \mathbb{D}_{1+\epsilon}$ denotes the oppositive sense of
$\partial \mathbb{D}_{1+\epsilon}$. $L$  divides the extended
complex plane into two parts $S^{-}=S^{+}_{\sharp2}\cup
S^{-}_{\sharp1}$ and $S^{+}=S^{-}_{\sharp2}\cup S^{+}_{\sharp1}$, or
simply,
\begin{equation} S^-=\{z:\,
|z|>1+\epsilon\,\,\mathrm{or}\,\,|z|<1-\epsilon\},\hspace{4mm}
S^+=\{z:\, 1-\epsilon<|z|<1+\epsilon\}.
\end{equation} Then, $\mathbf{R}$ is the solution of the following MRP:
\begin{equation}\left\{
\begin{array}{ll}
\mathbf{R}^+(t)=\mathbf{V}(t)\mathbf{R}^{-}(t),\hspace{3mm}&t\in L,\\[3mm]
\mathbf{P.P}[\mathbf{R},\infty](z)=I,&
\end{array}
\right.
\end{equation} where the factor matrix

\begin{equation} \mathbf{V}(t)=\left\{
\begin{array}{ll}
\left(\begin{array}{cc}
    1 &\displaystyle\frac{t^{-2n+1}(D^{-})^{2}(t)}{w(t)} \\[4mm]
    0 & 1
  \end{array}
  \right),\hspace{3mm}&t\in \partial \mathbb{D}_{1+\epsilon},\\[14mm]
   \left(
    \begin{array}{cc}
  1 &0\\[4mm]
 \displaystyle\frac{t^{2n-1}(D^+)^2(t)}{w(t)}& 1
  \end{array}
  \right),\hspace{3mm}&t\in\partial \mathbb{D}_{1-\epsilon}.
  \end{array}
  \right.
\end{equation}

\begin{rem}{\rm The MRP (3.53) is equivalent to the MRP (3.48) under the
transform (3.49), so it is also uniquely solvable.}
\end{rem}

\begin{rem}{\rm Since $w$ is analytic in some open neighborhood of $\partial \mathbb{D}$, for example,
the above $U(\epsilon)$, so is $V$,  say the annular region
\begin{equation}
\Omega(\epsilon)=\Omega_1(\epsilon)\cup\Omega_2(\epsilon)\hspace{3mm}
\end{equation} with
\begin{equation}
\Omega_1(\epsilon)= \left\{z:\,
1-\frac{5}{2}\epsilon<|z|<1-\frac{3}{2}\epsilon\right\}
\hspace{3mm}\mathrm{and}\hspace{3mm}\Omega_2(\epsilon)=\left\{z:
1+\frac{3}{2}\epsilon<|z|<1+\frac{5}{2}\epsilon\right\}.
\end{equation}
In this case, it is easy to see that there exist positive real
constants $C_{1}$ and $C_{2}$ such that
\begin{equation}
\|\mathbf{V}-I\|_{\Omega(\epsilon)}\leq C_1\exp\{-C_2(2n-1)\}.
\end{equation}
In \cite{dd}, they are given by
\begin{equation}
C_1=2 \max\left\{\left\|\frac{(D^+)^2}{w}\right\|_{\partial
\mathbb{D}},\,\,\left\|\frac{(D^-)^2}{w}\right\|_{\partial
\mathbb{D}}\right\} \hspace{2mm}\mathrm{and}\hspace{2mm}
C_2=\log\left(1+\frac{3}{2}\epsilon\right).
\end{equation}}
\end{rem}

\section{Asymptotic Analysis for Orthogonal Trigonometric Polynomials}

Tracing back the steps
$\mathbf{Y}\mapsto\mathbf{D}\mapsto\mathbf{F}\mapsto\mathbf{S}\mapsto\mathbf{R}$,
 we can obtain some asymptotic formulae for the first orthogonal
Laurent polynomials on the unit circle with respect to the positive
and analytic weight $w$. Consequently, we also obtain asymptotics
for the first orthogonal trigonometric polynomials.

{\bf Case 1:} $z\in S^{+}_{\sharp2}$.
 In this case,   we have
 \begin{equation}\begin{array}{lll}
\mathbf{Y}(z)&=&\left(\begin{array}{cc}
z^{2n}D^{-1}(z)&0\\[1mm]
0&z^{-2n+1}D(z)\end{array}\right)\mathbf{D}(z)\\[5mm]
&=&\left(\begin{array}{cc}
z^{2n}D^{-1}(z)&0\\[1mm]
0&z^{-2n+1}D(z)\end{array}\right)\mathbf{F}(z)\\[5mm]
&=&\left(\begin{array}{cc}
z^{2n}D^{-1}(z)&0\\[1mm]
0&z^{-2n+1}D(z)\end{array}\right)\mathbf{S}(z)\\[5mm]
&=&\left(\begin{array}{cc}
z^{2n}D^{-1}(z)&0\\[1mm]
0&z^{-2n+1}D(z)\end{array}\right)\mathbf{R}(z).
\end{array}
 \end{equation}
Let
\begin{equation} \mathbf{R}(z)=\left(
\begin{array}{cc}
R_{1,1}(z)&R_{1,2}(z)\\[1mm]
R_{2,1}(z)&R_{2,2}(z)
\end{array}
\right),
\end{equation} then, from (3.3), we have
\begin{equation}
a_{n}\sigma_{n}(z)+(\beta_{n}+i)b_{n}\pi_{n}(z)=\frac{z^n}{D(z)}R_{1,1}(z)
\end{equation}
and
\begin{equation}
\frac{1}{2}[a^{-1}_{n-1}(1+\beta_{n-1}i)\sigma_{n-1}(z)-ib^{-1}_{n-1}\pi_{n-1}(z)]=-\frac{z^n}{D(z)}R_{1,2}(z).
\end{equation} Now, quoting (3.56), Lemma 2.4, (4.3) and (4.4), we get
the estimations
\begin{align}
&\left|z^{-n}D(z)[a_{n}\sigma_{n}(z)+(\beta_{n}+i)b_{n}\pi_{n}(z)]-1\right|_{z\in
S_{\sharp2}^+}\\
\leq&C_1\exp\left\{-C_2(2n-1)\right\}\nonumber
\end{align} and
\begin{align}
&\left|\frac{1}{2}z^{-n}D(z)[a^{-1}_{n-1}(1+\beta_{n-1}i)\sigma_{n-1}(z)-ib^{-1}_{n-1}\pi_{n-1}(z)]\right|_{z\in
S_{\sharp2}^+}\\
\leq&C_1\exp\left\{-C_2(2n-1)\right\}\nonumber
\end{align} for sufficiently large $n$ with the constants $C_1$ and $C_2$ given in (3.57).

{\bf Case 2:} $z\in S^{-}_{\sharp1}$.
 In exactly the same way, we have
\begin{equation}
\begin{array}{lll} \mathbf{Y}(z)&=&\left(\begin{array}{cc}
D^{-1}(z)&0\\[2mm]
0&D(z)\end{array}\right)\mathbf{D}(z)\\[5mm]
&=&\left(\begin{array}{cc}
D^{-1}(z)&0\\[2mm]
0&D(z)\end{array}\right)\left(\begin{array}{cc}
z&0\\[2mm]
0&1\end{array}\right)\mathbf{F}(z)\\[5mm]
&=&\left(\begin{array}{cc}
D^{-1}(z)&0\\[2mm]
0&D(z)\end{array}\right)\left(\begin{array}{cc}
z&0\\[2mm]
0&1\end{array}\right)\mathbf{S}(z)\\[5mm]
&=&\left(\begin{array}{cc}
D^{-1}(z)&0\\[2mm]
0&D(z)\end{array}\right)\left(\begin{array}{cc}
z&0\\[2mm]
0&1\end{array}\right) \left(\begin{array}{cc}
0&-1\\[1mm]
1&0\end{array}\right) \mathbf{R}(z).
\end{array}
\end{equation}
Then, we have
\begin{equation}
 a_{n}\sigma_{n}(z)+(\beta_{n}+i)b_{n}\pi_{n}(z)=-\frac{1}{z^{n-1}D(z)}R_{2,1}(z)
\end{equation}
and
\begin{equation}
 \frac{1}{2}[a^{-1}_{n-1}(1+\beta_{n-1}i)\sigma_{n-1}(z)-ib^{-1}_{n-1}\pi_{n-1}(z)]=\frac{1}{z^{n-1}D(z)}R_{2,2}(z)
\end{equation}
for $z\in S^{-}_{\sharp1}\setminus\{0\}$. Now, quoting (3.56), Lemma
2.4, (4.8) and (4.9), we also get the estimations

\begin{equation}
\left|z^{n-1}D(z)[a_{n}\sigma_{n}(z)+(\beta_{n}+i)b_{n}\pi_{n}(z)]\right|_{z\in
S_{\sharp1}^-}\leq C_1\exp\left\{-C_2(2n-1)\right\}
\end{equation}
and
\begin{align}
&\left|\frac{1}{2}z^{n-1}D(z)[a^{-1}_{n-1}(1+\beta_{n-1}i)\sigma_{n-1}(z)-ib^{-1}_{n-1}\pi_{n-1}(z)]-1\right|_{z\in
S_{\sharp1}^-}\\
\leq&C_1\exp\left\{-C_2(2n-1)\right\}\nonumber
\end{align} for sufficiently large $n$ with the constants $C_1$ and
$C_2$ given in (3.57).

{\bf Case 3:} $z\in S^{-}_{\sharp2}$.
 In this case,   we have
\begin{equation}
\begin{array}{ll} \,\,\,\mathbf{Y}(z)\!\!\!
&=\left(\!\!\begin{array}{cc}
z^{2n}D^{-1}(z)&0\\[2mm]
0&z^{-2n+1}D(z)\end{array}\right)\mathbf{D}(z)\\[5mm]
&=\left(\!\!\begin{array}{cc}
z^{2n}D^{-1}(z)&0\\[2mm]
0&z^{-2n+1}D(z)\end{array}\right)\mathbf{F}(z)\\[5mm]
&=\left(\!\!\begin{array}{cc}
z^{2n}D^{-1}(z)&0\\[2mm]
0&z^{-2n+1}D(z)\end{array}\!\!\right)\!\!
\left(\!\!\begin{array}{cc}
1&-\displaystyle\frac{z^{-2n+1}D^{2}(z)}{w(z)}\\[3mm]
0&1
\end{array}
\!\!\right)\! \mathbf{S}(z)\\[5mm]
&=\left(\!\!\begin{array}{cc}
z^{2n}D^{-1}(z)&0\\[2mm]
0&z^{-2n+1}D(z)\end{array}\!\!\right)\!\!
\left(\!\!\begin{array}{cc}
1&-\displaystyle\frac{z^{-2n+1}D^{2}(z)}{w(z)}\\[3mm]
0&1
\end{array}
\!\!\right)\! \mathbf{R}(z).
\end{array}
\end{equation}
Then, we have
\begin{equation}
a_{n}\sigma_{n}(z)+(\beta_{n}+i)b_{n}\pi_{n}(z)=\frac{z^n}{D(z)}R_{1,1}(z)-\frac{D(z)}{z^{n-1}w(z)}R_{2,1}(z)
\end{equation}
and
\begin{align}
&\frac{1}{2}[a^{-1}_{n-1}(1+\beta_{n-1}i)\sigma_{n-1}(z)-ib^{-1}_{n-1}\pi_{n-1}(z)]\\
=&-\frac{z^n}{D(z)}R_{1,2}(z)+\frac{D(z)}{z^{n-1}w(z)}R_{2,2}(z).\nonumber
\end{align}

Again, by (3.56), Lemma 2.4, (4.13) and (4.14), we get the
estimations

\begin{align}
&\left|z^{-n}D(z)[a_{n}\sigma_{n}(z)+(\beta_{n}+i)b_{n}\pi_{n}(z)]-1\right|_{z\in
S_{\sharp2}^-} \\
\leq& C_1\left[1+\left\|\displaystyle\frac{D^2}{w}\!\right\|_{z\in
S^{-}_{\sharp2}}\right]\exp\left\{-C_2(2n-1)\right\}\nonumber
\end{align}
and
\begin{align}
&\left|\frac{1}{2}z^{n-1}D^{-1}(z)w(z)[a^{-1}_{n-1}(1+\beta_{n-1}i)\sigma_{n-1}(z)-ib^{-1}_{n-1}\pi_{n-1}(z)]\!-\!1\right|_{z\in
S_{\sharp2}^-}\!\! \\
\leq
&C_1\left[1\!+\!(1+\epsilon)^{2n-1}\left\|\displaystyle\frac{w}{D^2}\!\right\|_{z\in
S^{-}_{\sharp2}}\right]\!\exp\left\{\!-\!C_2(2n-1)\right\}\nonumber
\end{align} for sufficiently large $n$ with the constants
$C_1$ and $C_2$ given in (3.57).

{\bf Case 4:} $z\in S^{+}_{\sharp1}$.
 In this case,   we have
\begin{equation}
\begin{array}{lll} \mathbf{Y}(z)&=&\left(\begin{array}{cc}
D^{-1}(z)&0\\[2mm]
0&D(z)\end{array}\right)\mathbf{D}(z)\\[5mm]
&=&\left(\begin{array}{cc}
D^{-1}(z)&0\\[2mm]
0&D(z)\end{array}\right)\left(\begin{array}{cc}
z&0\\[2mm]
0&1\end{array}\right)\mathbf{F}(z)\\[5mm]
&=&\left(\begin{array}{cc}
zD^{-1}(z)&0\\[2mm]
0&D(z)\end{array}\right)\left(\begin{array}{cc}
    1 &\displaystyle\frac{z^{2n-1}D^2(z)}{w(z)} \\[3mm]
    0 & 1
  \end{array}
  \right)\mathbf{S}(z)\\[5mm]
&=&\left(\begin{array}{cc}
zD^{-1}(z)&0\\[2mm]
0&D(z)\end{array}\right)\left(\begin{array}{cc}
    1 &\displaystyle\frac{z^{2n-1}D^2(z)}{w(z)} \\[3mm]
    0 & 1
  \end{array}
  \right) \left(\begin{array}{cc}
0&-1\\[1mm]
1&0\end{array}\right) \mathbf{R}(z).
\end{array}
\end{equation}
Then, we have
\begin{equation}
a_{n}\sigma_{n}(z)+(\beta_{n}+i)b_{n}\pi_{n}(z)=\frac{z^nD(z)}{w(z)}R_{1,1}(z)-\frac{1}{z^{n-1}D(z)}R_{2,1}(z)
\end{equation}
and
\begin{align}
&\frac{1}{2}[a^{-1}_{n-1}(1+\beta_{n-1}i)\sigma_{n-1}(z)-ib^{-1}_{n-1}\pi_{n-1}(z)]\\
=&-\frac{z^nD(z)}{w(z)}R_{1,2}(z)+\frac{1}{z^{n-1}D(z)}R_{2,2}(z).\nonumber
\end{align}

All and the same, by (3.56), Lemma 2.4, (4.19) and (4.20), we get
the estimations

\begin{align}
&\left|z^{-n}D^{-1}(z)w(z)[a_{n}\sigma_{n}(z)+(\beta_{n}+i)b_{n}\pi_{n}(z)]-1\right|_{z\in
S_{\sharp1}^+} \\
\leq&
C_1\left[1+(1-\epsilon)^{-2n+1}\left\|\displaystyle\frac{w}{D^2}\!\right\|_{z\in
S^{+}_{\sharp1}}\right]\exp\left\{-C_2(2n-1)\right\}\nonumber
\end{align}
and
\begin{align}
&\left|\frac{1}{2}z^{n-1}D(z)[a^{-1}_{n-1}(1+\beta_{n-1}i)\sigma_{n-1}(z)-ib^{-1}_{n-1}\pi_{n-1}(z)]\!-\!1\right|_{z\in
S_{\sharp1}^+}\!\! \\
\leq
&C_1\left[1\!+\!\left\|\displaystyle\frac{D^2}{w}\!\right\|_{z\in
S^{+}_{\sharp1}}\right]\!\exp\left\{\!-\!C_2(2n-1)\right\}\nonumber
\end{align} for sufficiently large $n$ with the constants
$C_1$ and $C_2$ given in (3.57).

Now we can give the asymptotics for the first orthonormal Laurent
and trigonometric polynomials on the unit circle with respect to a
positive and analytic weight $w$.

\begin{thm}
Let $w$ be a positive and analytic weight function on the unit
circle $\partial \mathbb{D}$ and
\begin{equation}D(z)=\exp\left\{-\frac{1}{2\pi
i}\int_{\partial \mathbb{D}}\frac{\log
w(\tau)}{\tau-z}d\tau\right\},\hspace{4mm}|z|>1.
\end{equation} If
$\left\{1, \pi_{n}, \sigma_n\right\}$ denotes the system of the
first orthonormal Laurent polynomials on the unit circle with
respect to $w$, then
\begin{align}
a_{n}\sigma_{n}(z)=-\Lambda_{n}^{-1}\left[\frac{iz^n}{2b^{2}_{n}D(z)}R^{(n)}_{1,1}(z)-
                  \frac{(\beta_{n}+i)z^{n+1}}{D(z)}R^{(n+1)}_{1,2}(z)\right]
\end{align}
and
\begin{align}
b_{n}\pi_{n}(z)=-\Lambda_{n}^{-1}\left[\frac{(1+\beta_{n}i)z^n}{2a^{2}_{n}D(z)}R^{(n)}_{1,1}(z)+
                  \frac{z^{n+1}}{D(z)}R^{(n+1)}_{1,2}(z)\right]
\end{align}
 uniformly hold for $|z|\geq R$ with any $R>1$, where
 $\Lambda_{n}=-\frac{1}{2}[a_{n}^{-2}(1+\beta_{n}^{2})+b_{n}^{-2}]i\neq0$,
 $R^{(n)}_{1,1}(z)=1+\mathcal{O}\big(e^{-C_{2}{(2n-1)}}\big)$,
 $R^{(n+1)}_{1,2}(z)=\mathcal{O}\big(e^{-C_{2}{(2n+1)}}\big)$
with $C_2=\log\left(1+\frac{3}{2}\epsilon\right)$ in which
$0<\epsilon<\min\{\frac{1}{3}, R-1\}$, and $a_{n},b_{n},\beta_{n}$
are given in (2.17). Moreover,

\begin{equation}
\begin{cases}
a_{n}^{-1}z^{-n-1}\sigma_{n}(z)\sim-\frac{1}{2}\Lambda_{n}^{-1}a_{n}^{-2}b_{n}^{-2}iz^{-1}D^{-1}(z)
\,\,\,\,\mathrm{as}\,\,\,\,n\rightarrow\infty,\,\,\,\mathrm{when}\,\, a_{n}\geq 1,\\[2mm]
a_{n}z^{-n-1}\sigma_{n}(z)\sim-\frac{1}{2}\Lambda_{n}^{-1}b_{n}^{-2}iz^{-1}D^{-1}(z)
\,\,\,\,\mathrm{as}\,\,\,\,n\rightarrow\infty,\,\,\,\mathrm{when}\,\,
0<a_{n}<1
\end{cases}
\end{equation}
and
\begin{equation}
\begin{cases}
b^{-1}_{n}z^{-n-1}\pi_{n}(z)\sim-\frac{1}{2}\Lambda_{n}^{-1}a_{n}^{-2}b_{n}^{-2}(1+\beta_{n}i)z^{-1}D^{-1}(z)
\,\,\,\,\mathrm{as}\,\,\,\,n\rightarrow\infty,\,\,\,\mathrm{when}\,\, b_{n}\geq 1,\\[2mm]
b_{n}z^{-n-1}\pi_{n}(z)\sim-\frac{1}{2}\Lambda_{n}^{-1}a_{n}^{-2}(1+\beta_{n}i)z^{-1}D^{-1}(z)
\,\,\,\,\mathrm{as}\,\,\,\,n\rightarrow\infty,\,\,\,\mathrm{when}\,\,
0<b_{n}<1
\end{cases}
\end{equation}
uniformly hold for $|z|\geq R$ with any $R>1$. Here the asymptotic
symbol $\sim$ is defined by $f\sim g$ if and only if
$(f-g)\rightarrow 0$.
\end{thm}

\begin{proof} Since all the steepest descent transforms
$\mathbf{Y}\mapsto\mathbf{D}\mapsto\mathbf{F}\mapsto\mathbf{S}\mapsto\mathbf{R}$
depend on $n$, now we explicitly replace $\mathbf{R}$ by
$\mathbf{R}^{(n)}$ to indicate the $n$-dependence. Choose a small
enough $\epsilon$ so that $\{z:\,|z|\geq R\}\subset S^+_{\sharp2}$.
Thus, by (4.3) and (4.4), we get
\begin{equation}
a_{n}\sigma_{n}(z)+(\beta_{n}+i)b_{n}\pi_{n}(z)=\frac{z^n}{D(z)}R^{(n)}_{1,1}(z)
\end{equation}
and
\begin{equation}
\frac{1}{2}[a^{-1}_{n}(1+\beta_{n}i)\sigma_{n}(z)-ib^{-1}_{n}\pi_{n}(z)]=-\frac{z^{n+1}}{D(z)}R^{(n+1)}_{1,2}(z).
\end{equation}

Noting Remark 3.2, from (4.27) and (4.28), we have
\begin{align}
a_{n}\sigma_{n}(z)&=\Lambda_{n}^{-1}\begin{vmatrix}
                                    z^nD^{-1}(z)R^{(n)}_{1,1}(z) & \beta_{n}+i
                                     \\[2mm]
                                     -z^{n+1}D^{-1}(z)R^{(n+1)}_{1,2}(z) & -\frac{1}{2}b^{-2}_{n}i \\
                                   \end{vmatrix}\\[3mm]
                  &=-\Lambda_{n}^{-1}\left[\frac{iz^n}{2b^{2}_{n}D(z)}R^{(n)}_{1,1}(z)-
                  \frac{(\beta_{n}+i)z^{n+1}}{D(z)}R^{(n+1)}_{1,2}(z)\right]\nonumber
\end{align}
and
\begin{align}
b_{n}\pi_{n}(z)&=\Lambda_{n}^{-1}\begin{vmatrix}
                                  1&  z^nD^{-1}(z)R^{(n)}_{1,1}(z)
                                     \\[2mm]
                                  \frac{1}{2}a^{-2}_{n}(1+\beta_{n}i) &  -z^{n+1}D^{-1}(z)R^{(n+1)}_{1,2}(z) \\
                                   \end{vmatrix}\\[3mm]
                  &=-\Lambda_{n}^{-1}\left[\frac{(1+\beta_{n}i)z^n}{2a^{2}_{n}D(z)}R^{(n)}_{1,1}(z)+
                  \frac{z^{n+1}}{D(z)}R^{(n+1)}_{1,2}(z)\right],\nonumber
\end{align}
where $\Lambda_{n}$ is given in (3.34). By (3.56), we know that
\begin{equation}
R^{(n)}_{1,1}(z)=1+\mathcal{O}\big(e^{-C_{2}{(2n-1)}}\big)\,\,\,\,\mathrm{and}\,\,\,\,
R^{(n+1)}_{1,2}(z)=\mathcal{O}\big(e^{-C_{2}{(2n+1)}}\big)
\end{equation}
uniformly hold for $|z|\geq R$ with any $R>1$, where $C_{2}$ is
given in (3.57) with $0<\epsilon<\min\{\frac{1}{3},R-1\}$.

Now we turn to (4.25) and (4.26) and only verify (4.25). (4.26)
follows similarly. The verification is divided into two cases.

Case I: $a_{n}\geq 1$. In this case, by (4.29),
\begin{align}
&\left|a_{n}^{-1}z^{-n-1}D(z)\sigma_{n}(z)+\frac{1}{2}\Lambda_{n}^{-1}a_{n}^{-2}b_{n}^{-2}iz^{-1}\right|\\
\leq&C_{1}\left[\frac{1}{2}|\Lambda_{n}|^{-1}b_{n}^{-2}a_{n}^{-2}|z|^{-1}+|\Lambda_{n}|^{-1}a_{n}^{-2}\sqrt{1+\beta_{n}^{2}}
\right]\exp\{-C_{2}(2n-1)\}\nonumber\\
<&C_{1}(2+1/R)\exp\{-C_{2}(2n-1)\}\nonumber
\end{align}
uniformly holds for $|z|\geq R$ with any $R>1$ since
$\frac{1}{2}|\Lambda_{n}|^{-1}b_{n}^{-2}<1$,
$\frac{1}{2}|\Lambda_{n}|^{-1}a_{n}^{-2}(1+\beta_{n}^{2})<1$ and
$a^{-1}_{n}\leq 1$, where $C_{1}$ and $C_{2}$ are given in (3.57)
with $0<\epsilon<\min\{\frac{1}{3},R-1\}$.

Case II. $0<a_{n}<1$. Again, by (4.29),
\begin{align}
&\left|a_{n}z^{-n-1}D(z)\sigma_{n}(z)+\frac{1}{2}\Lambda_{n}^{-1}b_{n}^{-2}iz^{-1}\right|\\
\leq&C_{1}\left[\frac{1}{2}|\Lambda_{n}|^{-1}b_{n}^{-2}|z|^{-1}+|\Lambda_{n}|^{-1}
\big(a_{n}^{-2}\sqrt{1+\beta_{n}^{2}}\,\big)a_{n}^{2}
\right]\exp\{-C_{2}(2n-1)\}\nonumber\\
<&C_{1}(2+1/R)\exp\{-C_{2}(2n-1)\}\nonumber
\end{align}
uniformly holds for $|z|\geq R$ with any $R>1$ since
$\frac{1}{2}|\Lambda_{n}|^{-1}b_{n}^{-2}<1$,
$\frac{1}{2}|\Lambda_{n}|^{-1}a_{n}^{-2}(1+\beta_{n}^{2})<1$ and
$0<a_{n}<1$, where $C_{1}$ and $C_{2}$ are given in (3.57) with
$0<\epsilon<\min\{\frac{1}{3},R-1\}$.

So (4.25) follows from (4.32) and (4.33).
\end{proof}

\begin{thm}
Let $w$ be a positive and analytic weight function on the unit
circle $\partial \mathbb{D}$ and
\begin{equation}D(z)=\exp\left\{\frac{1}{2\pi
i}\int_{\partial \mathbb{D}}\frac{\log
w(\tau)}{\tau-z}d\tau\right\},\hspace{4mm}|z|<1.
\end{equation} If
$\left\{1, \pi_{n}, \sigma_n\right\}$ denotes the system of the
first orthonormal Laurent polynomials on the unit circle with
respect to $w$, then
\begin{align}
a_{n}\sigma_{n}(z)=\Lambda_{n}^{-1}\left[\frac{i}{2b^{2}_{n}z^{n-1}D(z)}R^{(n)}_{2,1}(z)-
                  \frac{(\beta_{n}+i)}{z^{n}D(z)}R^{(n+1)}_{2,2}(z)\right]
\end{align}
and
\begin{align}
b_{n}\pi_{n}(z)=\Lambda_{n}^{-1}\left[\frac{(1+\beta_{n}i)}{2a^{2}_{n}z^{n-1}D(z)}R^{(n)}_{2,1}(z)+
                  \frac{1}{z^{n}D(z)}R^{(n+1)}_{2,2}(z)\right]
\end{align}
 uniformly hold for $0<|z|\leq r$ with any $0<r<1$, where
 $\Lambda_{n}=-\frac{1}{2}[a_{n}^{-2}(1+\beta_{n}^{2})+b_{n}^{-2}]i\neq0$,
 $R^{(n)}_{2,1}(z)=\mathcal{O}\big(e^{-C_{2}{(2n-1)}}\big)$,
 $R^{(n+1)}_{2,2}(z)=1+\mathcal{O}\big(e^{-C_{2}{(2n+1)}}\big)$
with $C_2=\log\left(1+\frac{3}{2}\epsilon\right)$ in which
$0<\epsilon<\min\{\frac{1}{3}, 1-r\}$, and $a_{n},b_{n},\beta_{n}$
are given in (2.17). In addition,
\begin{equation}
\begin{cases}
a_{n}^{-1}z^{n}\sigma_{n}(z)\sim-\Lambda_{n}^{-1}a_{n}^{-2}(\beta_{n}+i)D^{-1}(z)
\,\,\,\,\mathrm{as}\,\,\,\,n\rightarrow\infty,\,\,\,\mathrm{when}\,\, a_{n}\geq 1,\\[2mm]
a_{n}z^{n}\sigma_{n}(z)\sim-\Lambda_{n}^{-1}(\beta_{n}+i)D^{-1}(z)
\,\,\,\,\mathrm{as}\,\,\,\,n\rightarrow\infty,\,\,\,\mathrm{when}\,\,
0<a_{n}<1
\end{cases}
\end{equation}
and
\begin{equation}
\begin{cases}
b^{-1}_{n}z^{n}\pi_{n}(z)\sim\Lambda_{n}^{-1}b_{n}^{-2}D^{-1}(z)
\,\,\,\,\mathrm{as}\,\,\,\,n\rightarrow\infty,\,\,\,\mathrm{when}\,\, b_{n}\geq 1,\\[2mm]
b_{n}z^{n}\pi_{n}(z)\sim\Lambda_{n}^{-1}D^{-1}(z)
\,\,\,\,\mathrm{as}\,\,\,\,n\rightarrow\infty,\,\,\,\mathrm{when}\,\,
0<b_{n}<1
\end{cases}
\end{equation}
uniformly hold for $0<|z|\leq r$ with any $0<r<1$. Here the
asymptotic symbol $\sim$ is defined by $f\sim g$ if and only if
$(f-g)\rightarrow 0$.
\end{thm}

\begin{proof}
Pick a small enough $\epsilon$ so that $\{z:\,0<|z|\leq r\}\subset
S^-_{\sharp1}$. By (4.8) and (4.9),
\begin{equation}
 a_{n}\sigma_{n}(z)+(\beta_{n}+i)b_{n}\pi_{n}(z)=-\frac{1}{z^{n-1}D(z)}R^{(n)}_{2,1}(z),
\end{equation}
and
\begin{equation}
 \frac{1}{2}[a^{-1}_{n}(1+\beta_{n}i)\sigma_{n}(z)-ib^{-1}_{n}\pi_{n}(z)]=
 \frac{1}{z^{n}D(z)}R^{(n+1)}_{2,2}(z).
\end{equation}
Then,
\begin{align}
a_{n}\sigma_{n}(z)&=\Lambda_{n}^{-1}\begin{vmatrix}
                                    -z^{-n+1}D^{-1}(z)R^{(n)}_{2,1}(z) & \beta_{n}+i
                                     \\[2mm]
                                     z^{-n}D^{-1}(z)R^{(n+1)}_{2,2}(z) & -\frac{1}{2}b^{-2}_{n}i \\
                                   \end{vmatrix}\\[3mm]
                  &=\Lambda_{n}^{-1}\left[\frac{i}{2b^{2}_{n}z^{n-1}D(z)}R^{(n)}_{2,1}(z)-
                  \frac{(\beta_{n}+i)}{z^{n}D(z)}R^{(n+1)}_{2,2}(z)\right]\nonumber
\end{align}
and
\begin{align}
b_{n}\pi_{n}(z)&=\Lambda_{n}^{-1}\begin{vmatrix}
                                  1&  -z^{-n+1}D^{-1}(z)R^{(n)}_{2,1}(z)
                                     \\[2mm]
                                  \frac{1}{2}a^{-2}_{n}(1+\beta_{n}i) &  z^{-n}D^{-1}(z)R^{(n+1)}_{2,2}(z) \\
                                   \end{vmatrix}\\[3mm]
                  &=\Lambda_{n}^{-1}\left[\frac{(1+\beta_{n}i)}{2a^{2}_{n}z^{n-1}D(z)}R^{(n)}_{2,1}(z)+
                  \frac{1}{z^{n}D(z)}R^{(n+1)}_{2,2}(z)\right],\nonumber
\end{align}
where $\Lambda_{n}$ is given in (3.34). Again, by (3.56), we know
\begin{equation}
R^{(n)}_{2,1}(z)=\mathcal{O}\big(e^{-C_{2}{(2n-1)}}\big)\,\,\,\,\mathrm{and}\,\,\,\,
R^{(n+1)}_{2,2}(z)=1+\mathcal{O}\big(e^{-C_{2}{(2n+1)}}\big)
\end{equation}
 uniformly hold for $0<|z|\leq r$ with any $0<r<1$, where $C_{2}$ is given in (3.57) with
 $0<\epsilon<\min\{\frac{1}{3}, 1-r\}$.

Next we consider (4.37) and (4.38) and only verify (4.37). (4.38)
follows similarly. It is also divided into two cases.

Case I: $a_{n}\geq 1$. In this case, by (4.41),
\begin{align}
&\left|a_{n}^{-1}z^{n}D(z)\sigma_{n}(z)+\Lambda_{n}^{-1}a_{n}^{-2}(\beta_{n}+i)\right|\\
\leq&C_{1}\left[\frac{1}{2}|\Lambda_{n}|^{-1}b_{n}^{-2}a_{n}^{-2}|z|+|\Lambda_{n}|^{-1}a_{n}^{-2}\sqrt{1+\beta_{n}^{2}}
\right]\exp\{-C_{2}(2n-1)\}\nonumber\\
<&C_{1}(2+r)\exp\{-C_{2}(2n-1)\}\nonumber
\end{align}
uniformly holds for $0<|z|\leq r$ with any $0<r<1$ since
$\frac{1}{2}|\Lambda_{n}|^{-1}b_{n}^{-2}<1$,
$\frac{1}{2}|\Lambda_{n}|^{-1}a_{n}^{-2}(1+\beta_{n}^{2})<1$ and
$a^{-1}_{n}\leq 1$, where $C_{1}$ and $C_{2}$ are given in (3.57)
with $0<\epsilon<\min\{\frac{1}{3}, 1-r\}$.

Case II. $0<a_{n}<1$. In a similar way, by (4.41),
\begin{align}
&\left|a_{n}z^{n}D(z)\sigma_{n}(z)+\Lambda_{n}^{-1}(\beta_{n}+i)\right|\\
\leq&C_{1}\left[\frac{1}{2}|\Lambda_{n}|^{-1}b_{n}^{-2}|z|+|\Lambda_{n}|^{-1}
\big(a_{n}^{-2}\sqrt{1+\beta_{n}^{2}}\,\big)a_{n}^{2}
\right]\exp\{-C_{2}(2n-1)\}\nonumber\\
<&C_{1}(2+r)\exp\{-C_{2}(2n-1)\}\nonumber
\end{align}
uniformly holds for $0<|z|\leq r$ with any $0<r<1$ since
$\frac{1}{2}|\Lambda_{n}|^{-1}b_{n}^{-2}<1$,
$\frac{1}{2}|\Lambda_{n}|^{-1}a_{n}^{-2}(1+\beta_{n}^{2})<1$ and
$0<a_{n}<1$, where $C_{1}$ and $C_{2}$ are given in (3.57) with
 $0<\epsilon<\min\{\frac{1}{3}, 1-r\}$.

So (4.37) follows from (4.44) and (4.45).
\end{proof}

\begin{thm}[Asymptotics of Orthogonal Trigonometric Polynomials]
Let $w$ be a positive and analytic weight function on the unit
circle $\partial \mathbb{D}$ and
\begin{equation}D(t)=\exp\left\{\frac{1}{2}\log w(t)-\frac{1}{2\pi
i}\int_{\partial \mathbb{D}}\frac{\log
w(\tau)}{\tau-t}d\tau\right\},\hspace{4mm}|t|=1.
\end{equation} If
$\left\{1, \pi_{n}, \sigma_n\right\}$ denotes the system of the
first orthonormal trigonometric polynomials on the unit circle with
respect to $w$, then
\begin{align}
a_{n}\sigma_{n}(\theta)=&-\Lambda_{n}^{-1}\Big[\frac{1}{2}b^{-2}_{n}i
\Big(\frac{e^{in\theta}}{D(e^{i\theta})}R^{(n)}_{1,1}(\theta)-\frac{D(e^{i\theta})}{e^{i(n-1)\theta}
w(\theta)}R^{(n)}_{2,1}(\theta)\Big)\\[3mm]
                  &-(\beta_{n}+i)\Big(\frac{e^{i(n+1)\theta}}{D(e^{i\theta})}R^{(n+1)}_{1,2}(\theta)
                  -\frac{D(e^{i\theta})}{e^{in\theta}w(\theta)}R^{(n+1)}_{2,2}(\theta)\Big)\Big]\nonumber
\end{align}
and
\begin{align}
b_{n}\pi_{n}(\theta)=&-\Lambda_{n}^{-1}\Big[\frac{1}{2}a^{-2}_{n}(1+\beta_{n}i)
\Big(\frac{e^{in\theta}}{D(e^{i\theta})}R^{(n)}_{1,1}(\theta)-\frac{D(e^{i\theta})}{e^{i(n-1)\theta}
w(\theta)}R^{(n)}_{2,1}(\theta)\Big)\\[3mm]
                  &+\Big(\frac{e^{i(n+1)\theta}}{D(e^{i\theta})}R^{(n+1)}_{1,2}(\theta)
                  -\frac{D(e^{i\theta})}{e^{in\theta}w(\theta)}R^{(n+1)}_{2,2}(\theta)\Big)\Big]\nonumber
\end{align}
 uniformly hold for $\theta\in[0, 2\pi)$, where
 $\Lambda_{n}=-\frac{1}{2}[a_{n}^{-2}(1+\beta_{n}^{2})+b_{n}^{-2}]i\neq0$,
 $R^{(n)}_{1,1}(\theta)=1+\mathcal{O}\big(e^{-C_{2}{(2n-1)}}\big)$,
 $R^{(n)}_{1,2}(\theta)=\mathcal{O}\big(e^{-C_{2}{(2n-1)}}\big)$,
 $R^{(n+1)}_{2,1}(\theta)=\mathcal{O}\big(e^{-C_{2}{(2n+1)}}\big)$,
 $R^{(n+1)}_{2,2}(\theta)=1+\mathcal{O}\big(e^{-C_{2}{(2n+1)}}\big)$
with $C_2=\log\left(1+\frac{3}{2}\epsilon\right)$ in which
$0<\epsilon<\frac{1}{3}$, and $a_{n},b_{n},\beta_{n}$ are given in
(2.17). In addition,
\begin{equation}
\begin{cases}
a_{n}^{-1}e^{-i(n+1)\theta}\sigma_{n}(\theta)\sim-\Lambda_{n}^{-1}\Big[\frac{1}{2}a_{n}^{-2}b_{n}^{-2}ie^{-i\theta}D^{-1}(e^{i\theta})
\\\hspace{28mm}\,+a_{n}^{-2}(\beta_{n}+i)e^{-i(2n+1)\theta}D(e^{i\theta})w^{-1}(\theta)\Big]
\,\,\mathrm{as}\,\,n\rightarrow\infty,\,\mathrm{when}\,\, a_{n}\geq 1,\\[2mm]
a_{n}e^{-i(n+1)\theta}\sigma_{n}(\theta)\sim-\Lambda_{n}^{-1}\Big[\frac{1}{2}b_{n}^{-2}ie^{-i\theta}D^{-1}(e^{i\theta})
\\\hspace{27mm}+(\beta_{n}+i)e^{-i(2n+1)\theta}D(e^{i\theta})w^{-1}(\theta)\Big]
\,\,\mathrm{as}\,\,n\rightarrow\infty,\,\mathrm{when}\,\, 0<a_{n}<1
\end{cases}
\end{equation}
and
\begin{equation}
\begin{cases}
b_{n}^{-1}e^{-i(n+1)\theta}\pi_{n}(\theta)\sim-\Lambda_{n}^{-1}\Big[\frac{1}{2}a_{n}^{-2}b_{n}^{-2}
(1+\beta_{n}i)e^{-i\theta}D^{-1}(e^{i\theta})
\\\hspace{28mm}\,-b_{n}^{-2}e^{-i(2n+1)\theta}D(e^{i\theta})w^{-1}(\theta)\Big]
\,\,\mathrm{as}\,\,n\rightarrow\infty,\,\mathrm{when}\,\, b_{n}\geq 1,\\[2mm]
b_{n}e^{-i(n+1)\theta}\pi_{n}(\theta)\sim-\Lambda_{n}^{-1}\Big[\frac{1}{2}a_{n}^{-2}
(1+\beta_{n}i)e^{-i\theta}D^{-1}(e^{i\theta})
\\\hspace{26mm}\,-e^{-i(2n+1)\theta}D(e^{i\theta})w^{-1}(\theta)\Big]
\,\,\mathrm{as}\,\,n\rightarrow\infty,\,\mathrm{when}\,\, 0<b_{n}<1
\end{cases}
\end{equation}
uniformly hold for $\theta\in[0, 2\pi)$. Here the asymptotic symbol
$\sim$ is defined by $f\sim g$ if and only if $(f-g)\rightarrow 0$.
\end{thm}

\begin{proof}
For $z\in S^{-}_{\sharp2}$, by (4.13) and (4.14),
\begin{equation}
a_{n}\sigma_{n}(z)+(\beta_{n}+i)b_{n}\pi_{n}(z)=\frac{z^n}{D(z)}R^{(n)}_{1,1}(z)-\frac{D(z)}{z^{n-1}w(z)}R^{(n)}_{2,1}(z)
\end{equation}
and
\begin{align}
&\frac{1}{2}[a^{-1}_{n}(1+\beta_{n}i)\sigma_{n}(z)-ib^{-1}_{n}\pi_{n}(z)]\\
=&-\frac{z^{n+1}}{D(z)}R^{(n+1)}_{1,2}(z)+\frac{D(z)}{z^{n}w(z)}R^{(n+1)}_{2,2}(z).\nonumber
\end{align}
Then,
\begin{align}
a_{n}\sigma_{n}(z)=&-\Lambda_{n}^{-1}\Big[\frac{1}{2}b^{-2}_{n}i
\Big(\frac{z^n}{D(z)}R^{(n)}_{1,1}(z)-\frac{D(z)}{z^{n-1}w(z)}R^{(n)}_{2,1}(z)\Big)\\[3mm]
                  &-(\beta_{n}+i)\Big(\frac{z^{n+1}}{D(z)}R^{(n+1)}_{1,2}(z)
                  -\frac{D(z)}{z^{n}w(z)}R^{(n+1)}_{2,2}(z)\Big)\Big]\nonumber
\end{align}
and
\begin{align}
b_{n}\pi_{n}(z)=&-\Lambda_{n}^{-1}\Big[\frac{1}{2}a^{-2}_{n}(1+\beta_{n}i)
\Big(\frac{z^n}{D(z)}R^{(n)}_{1,1}(z)-\frac{D(z)}{z^{n-1}w(z)}R^{(n)}_{2,1}(z)\Big)\\[3mm]
                  &+\Big(\frac{z^{n+1}}{D(z)}R^{(n+1)}_{1,2}(z)
                  -\frac{D(z)}{z^{n}w(z)}R^{(n+1)}_{2,2}(z)\Big)\Big]\nonumber
\end{align}
where $\Lambda_{n}$ is given in (3.34). Again, by (3.56), we know
\begin{equation}
R^{(n)}_{1,1}(z)=1+\mathcal{O}\big(e^{-C_{2}{(2n-1)}}\big),\,\,\,\,\,\,
R^{(n)}_{2,1}(z)=\mathcal{O}\big(e^{-C_{2}{(2n-1)}}\big),
\end{equation}
and
\begin{equation}
R^{(n+1)}_{1,2}(z)=\mathcal{O}\big(e^{-C_{2}{(2n+1)}}\big),\,\,\,\,
R^{(n+1)}_{2,2}(z)=1+\mathcal{O}\big(e^{-C_{2}{(2n+1)}}\big)
\end{equation}
uniformly hold for $z\in S^{-}_{\sharp2}$, where $C_{2}$ is given in
(3.57) with $0<\epsilon<\frac{1}{3}$.

Let $t=e^{i\theta}$, $\theta \in [0, 2\pi)$. Take $z$ close to $t$,
$z\in S^{-}_{\sharp2}$, identifying $w(e^{i\theta})$ with
$w(\theta)$ via the map $[0,2\pi)\ni\theta\rightarrow e^{i\theta}\in
\partial \mathbb{D}$, it follows that (4.47) and (4.48) uniformly hold for
$\theta\in[0, 2\pi)$.

Next we turn to verify (4.49). (4.50) follows in an extremely
similar way. It is again divided into two cases.

Case I: $a_{n}\geq 1$. In this case, by (4.53),
\begin{align}
&\left|a_{n}^{-1}z^{-n-1}\sigma_{n}(z)+\Lambda_{n}^{-1}\Big[\frac{1}{2}b_{n}^{-2}a_{n}^{-2}iz^{-1}D^{-1}(z)
+a_{n}^{-2}(\beta_{n}+i)z^{-2n-1}D(z)w^{-1}(z)\Big]\right|\\
\leq&C_{1}\Big[\frac{1}{2}|\Lambda_{n}|^{-1}b_{n}^{-2}a_{n}^{-2}|z|^{-1}\Big(|D^{-1}(z)|+|z|^{-2n+1}|D(z)|
|w^{-1}(z)|\Big)\nonumber\\
&+|\Lambda_{n}|^{-1}a_{n}^{-2}\sqrt{1+\beta_{n}^{2}}\Big(|D^{-1}(z)|+|z|^{-2n-1}|D(z)|
|w^{-1}(z)|\Big)
\Big]\exp\{-C_{2}(2n-1)\}\nonumber\\
<&6C_{1}\left[\left\|\frac{1}{D}\right\|_{\partial
\mathbb{D}}+\left\|\frac{D}{w}\right\|_{\partial
\mathbb{D}}\right]\exp\{-C_{2}(2n-1)\}\nonumber
\end{align}
uniformly hold for $z\in S^{-}_{\sharp2}=\{z: 1<|z|<1+\epsilon\}$
with small enough $\epsilon>0$ since
$\frac{1}{2}|\Lambda_{n}|^{-1}b_{n}^{-2}<1$,
$\frac{1}{2}|\Lambda_{n}|^{-1}a_{n}^{-2}(1+\beta_{n}^{2})<1$ and
$a^{-1}_{n}\leq 1$, where $C_{1}$ and $C_{2}$ are given in (3.57).

Case II. $0<a_{n}<1$. In a similar way, by (4.53),
\begin{align}
&\left|a_{n}z^{-n-1}\sigma_{n}(z)+\Lambda_{n}^{-1}\Big[\frac{1}{2}b_{n}^{-2}iz^{-1}D^{-1}(z)
+(\beta_{n}+i)z^{-2n-1}D(z)w^{-1}(z)\Big]\right|\\
\leq&C_{1}\Big[\frac{1}{2}|\Lambda_{n}|^{-1}b_{n}^{-2}|z|^{-1}\Big(|D^{-1}(z)|+|z|^{-2n+1}|D(z)|
|w^{-1}(z)|\Big)\nonumber\\
&+|\Lambda_{n}|^{-1}\big(a_{n}^{-2}\sqrt{1+\beta_{n}^{2}}\,\big)a^{2}_{n}\Big(|D^{-1}(z)|+|z|^{-2n-1}|D(z)|
|w^{-1}(z)|\Big)
\Big]\exp\{-C_{2}(2n-1)\}\nonumber\\
<&6C_{1}\left[\left\|\frac{1}{D}\right\|_{\partial
\mathbb{D}}+\left\|\frac{D}{w}\right\|_{\partial
\mathbb{D}}\right]\exp\{-C_{2}(2n-1)\}\nonumber
\end{align}
uniformly hold for $z\in S^{-}_{\sharp2}=\{z: 1<|z|<1+\epsilon\}$
with small enough $\epsilon>0$ since
$\frac{1}{2}|\Lambda_{n}|^{-1}b_{n}^{-2}<1$,
$\frac{1}{2}|\Lambda_{n}|^{-1}a_{n}^{-2}(1+\beta_{n}^{2})<1$ and
$0<a_{n}<1$, where $C_{1}$ and $C_{2}$ are given in (3.57).

So (4.49) follows from (4.57) and (4.58) when $z\in S^{-}_{\sharp2}$
tends to $t=e^{i\theta}$, $\theta\in[0, 2\pi)$.
\end{proof}

\begin{rem}
Let
\begin{equation}D(t)=\exp\left\{\frac{1}{2}\log w(t)+\frac{1}{2\pi
i}\int_{\partial \mathbb{D}}\frac{\log
w(\tau)}{\tau-t}d\tau\right\},\hspace{4mm}|t|=1.
\end{equation}
By (4.18) and (4.19), we can also obtain the asymptotics for
orthogonal trigonometric polynomials on the unit circle with respect
to the positive and analytic weight $w$ in terms of a different
form. And it is easy to check that they are essentially equivalent.
\end{rem}

\section{Relations with Orthogonal Polynomials on the Unit Circle}

In this section, we establish the relationship between orthogonal
trigonometric polynomials (OTP) and orthogonal polynomials on the
unit circle (OPUC). That is, we find two transform formulae which
relate orthogonal trigonometric polynomials and orthogonal
polynomials on the unit circle to each other. By the transform
formulae and the theory of OPUC , we obtain some recurrent formulae,
Christoffel-Darboux formula and some properties of zeros for
orthogonal trigonometric polynomials on the unit cirlcle.
\subsection{OPUC Solution for Characterization of OTP}

In Section 3.1, we have given the characterization of orthogonal
trigonometric polynomials by a MRP (i.e., the MRP (3.1)) and
obtained its unique solution (3.3) in terms of orthogonal
trigonometric polynomials. In what follows, one will find that the
unique solution can be also expressed by orthogonal polynomials on
the unit circle. So we find out the relationship between orthogonal
trigonometric polynomials and orthogonal polynomials on the unit
circle by the uniqueness of solution of the characterization.

\begin{thm}
The MRP (3.1) is uniquely solvable and its unique solution is
\begin{equation}
\begin{pmatrix}
       z\Phi_{2n-1}(z) &
       -\kappa^{2}_{2n-2}z\Phi^{*}_{2n-2}(z)
       \\[2mm]
       \mathbf{C}\big[\tau\Phi_{2n-1}\big](z) &
       -\kappa^{2}_{2n-2}\mathbf{C}\big[\tau\Phi^{*}_{2n-2}\big](z)\\
     \end{pmatrix},
\end{equation}
where $\Phi_{2n-1}$ is the monic orthogonal polynomial of order
$2n-1$ on the unit circle with respect to the weight $w$,
$\kappa_{2n-2}$ is the leading coefficient of the orthonormal
polynomial of order $2n-2$ on the unit circle and
$\kappa_{2n-2}=\|\Phi_{2n-2}\|^{-1}_{\mathbb{C}}$, and
$\Phi^{*}_{2n-2}$ is the reversed polynomial of the monic orthogonal
polynomial $\Phi_{2n-2}$ of order $2n-2$ on the unit circle, and the
Cauchy singular integral operator $\mathbf{C}$ is given by (2.35).
\end{thm}

\begin{proof}
As in Section 3.1, the boundary value conditions in (3.1) can be
explicitly written as

\begin{align}
\begin{cases}
Y_{1,1}^+(t)=\,\,Y_{1,1}^{-}(t), \hspace{6mm}t\in \partial
\mathbb{D},\nonumber\\[2mm]
Y_{2,1}^+(t)=\,\,t^{-2n}w(t)Y_{1,1}^{-}(t)+Y_{2,1}^{-}(t), \hspace{6mm}t\in \partial \mathbb{D},\nonumber\\[2mm]
Y_{1,2}^+(t)=\,\,Y_{1,2}^{-}(t), \hspace{6mm}t\in \partial \mathbb{D},\nonumber\\[2mm]
Y_{2,2}^+(t)=\,\,t^{-2n}w(t)Y_{1,2}^{-}(t)+Y_{2,2}^{-}(t),
\hspace{6mm}t\in \partial \mathbb{D}\nonumber.
\end{cases}
\end{align}
The condition for the behavior of the principal part at infinity can
be clearly written as
\begin{equation*}
\left(\begin{array}{cc}
\mathbf{P.P}[z^{-2n}Y_{1,1},\infty](z)&\mathbf{P.P}[z^{-2n}Y_{1,2},\infty](z)\\[3mm]
\mathbf{P.P}[z^{2n-1}Y_{2,1},\infty](z)&\mathbf{P.P}[z^{2n-1}Y_{2,2},\infty](z)
\end{array}
\right)=\left(\begin{array}{cc}
1&0\\[3mm]
0&1
\end{array}
\right).
\end{equation*}
Obviously, $Y_{1,1}(z)$ is a monic polynomial of order $2n$. Note
that $Y_{1,1}(0)=0$, now we claim
\begin{equation}
Y_{1,1}(z)=zQ_{1,1}(z), \,\,\,\,\,Q_{1,1}\in \Pi_{2n-1},
\end{equation}
where $\Pi_{2n-1}$ denotes the set of all complex polynomials of
order at most $2n-1$, i.e., $Q_{1,1}$ is a monic polynomial of order
$2n-1$.

Noting the boundary value condition for $Y_{2,1}$, by (5.2), we have
\begin{align}
Y_{2,1}(z)&=\frac{1}{2\pi i}\int_{\partial
\mathbb{D}}\frac{\tau^{-2n+1}Q_{1,1}(\tau)w(\tau)}{\tau-z}d\tau\\
&=\frac{1}{2\pi i}\int_{\partial
\mathbb{D}}\frac{\tau^{-2n+2}Q_{1,1}(\tau)w(\tau)}{\tau-z}\frac{d\tau}{\tau}.
\end{align}
Since
\begin{equation*}
\frac{1}{\tau-z}=-\sum_{k=0}^{\infty}\frac{\tau^{k}}{z^{k+1}}
\end{equation*}
holds for sufficiently large $z$, by (5.4), from the condition at
infinity for $Y_{2,1}$, it is easy to see that
\begin{equation}
\frac{1}{2\pi i}\int_{\partial
\mathbb{D}}\tau^{-2n+2+k}Q_{1,1}(\tau)w(\tau)\frac{d\tau}{\tau}=0,\,\,k=0,1,\ldots,2n-2.
\end{equation}
This shows that
\begin{equation}
\langle\tau^{k},
Q_{1,1}\rangle_{\mathbb{C}}=0,\,\,\,\,k=0,1,\ldots,2n-2.
\end{equation}
Therefore,
\begin{equation}
Q_{1,1}(z)=\Phi_{2n-1}(z)
\end{equation}
is just the monic orthogonal polynomial of order $2n-1$ on the unit
circle. Hence,
\begin{equation}
Y_{1,1}(z)=z\Phi_{2n-1}(z)
\end{equation}
and
\begin{equation}
Y_{2,1}(z)=\frac{1}{2\pi i}\int_{\partial
\mathbb{D}}\frac{\tau\Phi_{2n-1}(\tau)w(\tau)}{\tau-z}d\tau.
\end{equation}

Similarly, $Y_{1,2}(z)$ is a polynomial of order $2n-1$. All and the
same, note that $Y_{1,2}(0)=0$, we assert
\begin{equation}
Y_{1,2}(z)=zQ_{1,2}(z),\,\,\,\,Q_{1,2}\in \Pi_{2n-2}.
\end{equation}
Thus we get
\begin{align}
Y_{2,2}(z)&=\frac{1}{2\pi i}\int_{\partial
\mathbb{D}}\frac{\tau^{-2n+1}Q_{1,2}(\tau)w(\tau)}{\tau-z}d\tau\\
&=\frac{1}{2\pi i}\int_{\partial
\mathbb{D}}\frac{\tau^{-2n+2}Q_{1,2}(\tau)w(\tau)}{\tau-z}\frac{d\tau}{\tau}.
\end{align}

By the condition at infinity for $Y_{2,2}$, we get
\begin{equation}
\frac{1}{2\pi i}\int_{\partial
\mathbb{D}}\tau^{-2n+2+k}Q_{1,2}(\tau)w(\tau)\frac{d\tau}{\tau}=0,\,\,k=0,1,\ldots,2n-3
\end{equation}
and
\begin{equation}
\frac{1}{2\pi i}\int_{\partial
\mathbb{D}}Q_{1,2}(\tau)w(\tau)\frac{d\tau}{\tau}=-1.
\end{equation}

(5.13) and (5.14) are equivalent to
\begin{equation}
\langle\tau^{k},
Q_{1,2}\rangle_{\mathbb{C}}=0,\,\,\,\,j=1,2,\ldots,2n-2
\end{equation}
and
\begin{equation}
\langle1, Q_{1,2}\rangle_{\mathbb{C}}=-1.
\end{equation}

From (5.15) and (5.16), we know that (see \cite{sim2})
\begin{equation}
Q_{1,2}(z)=-\kappa^{2}_{2n-2}\Phi^{*}_{2n-2}(z),
\end{equation}
where $\kappa_{2n-2}$ is the leading coefficient of the orthonormal
polynomial of order $2n-2$ on the unit circle with respect to the
weight $w$ and $\kappa_{2n-2}=\|\Phi_{2n-2}\|^{-1}_{\mathbb{C}}$,
and $\Phi^{*}_{2n-2}$ is just the reversed polynomial of the monic
orthogonal polynomial $\Phi_{2n-2}$ of order $2n-2$ on the unit
circle. So
\begin{equation}
Y_{1,2}(z)=-\kappa^{2}_{2n-2}z\Phi^{*}_{2n-2}(z)
\end{equation}
and
\begin{equation}
Y_{2,2}(z)=-\frac{\kappa^{2}_{2n-2}}{2\pi i}\int_{\partial
\mathbb{D}}\frac{\tau\Phi^{*}_{2n-2}(\tau)w(\tau)}{\tau-z}d\tau.
\end{equation}

By the uniqueness of OPUC, we complete the proof of Theorem 5.1.
\end{proof}

\subsection{Relations of OTP and OPUC}

Theorem 3.1 shows that the unique solution of the characterization
of orthogonal trigonometric polynomials can be expressed in terms of
orthogonal trigonometric polynomials. However, Theorem 5.1 shows
that the unique solution of the characterization of orthogonal
trigonometric polynomials can be also expressed in terms of
orthogonal polynomials on the unit circle. By the uniqueness of
solution, the relationship between OTP ans OPUC is implied in
Theorem 3.1 and 5.1.

In general, we have
\begin{thm} Let $\mu$ be a nontrivial probability measure on the
unit circle $\partial \mathbb{D}=\{z: |z|=1\}$, $\{1, \pi_{n},
\sigma_{n}\}$ be the unique system of the first orthonormal Laurent
polynomials on the unit circle with respect to $\mu$, and
$\{\Phi_{n}\}$ be the unique system of the monic orthogonal
polynomials on the unit circle with respect to $\mu$. Then for any
$z\in \mathbb{C}$ and $n\in \mathbb{N}$,
\begin{equation}
\Phi_{2n-1}(z)=z^{n-1}[a_{n}\sigma_{n}(z)+(\beta_{n}+i)b_{n}\pi_{n}(z)]
\end{equation}
and
\begin{equation}
\kappa^{2}_{2n}\Phi^{*}_{2n}(z)=\frac{1}{2}z^{n}[a^{-1}_{n}(1+\beta_{n}i)\sigma_{n}(z)
-ib^{-1}_{n}\pi_{n}(z)],
\end{equation}
where $\kappa_{2n}$ is the leading coefficient of the orthonormal
polynomial of order $2n$ on the unit circle with respect to $\mu$
and $\kappa_{2n}=\|\Phi_{2n}\|^{-1}_{\mathbb{C}}$, and $a_{n},
b_{n}, \beta_{n}$ are given in (2.17). By the convention in Section
2.1, (5.21) also holds when $n=0$.
\end{thm}

\begin{proof}
We only verify (5.20). (5.21) follows similarly. Let
\begin{equation}
Q(z)=z^{n-1}[a_{n}\sigma_{n}(z)+(\beta_{n}+i)b_{n}\pi_{n}(z)].
\end{equation}
By straightforward calculations, $Q\in \Pi_{2n-1}$ and it is easy to
see that its leading coefficient is 1 as well as
\begin{equation}
\langle\tau^{k}, Q\rangle_{\mathbb{C}}=0, \,\,\,\,k=0,1,\ldots,2n-2.
\end{equation}
Therefore,
\begin{equation}
Q(z)=\Phi_{2n-1}(z).
\end{equation}

So
\begin{equation*}
\Phi_{2n-1}(z)=z^{n-1}[a_{n}\sigma_{n}(z)+(\beta_{n}+i)b_{n}\pi_{n}(z)].\qedhere
\end{equation*}
\end{proof}

\begin{rem}
In fact, by (5.20) and (5.21), the estimations (4.5), (4.6), (4.10),
(4.11), (4.15), (4.16), (4.20) and (4.21) in Section 4 give the
asymptotics of $\Phi_{2n-1}(z)$ and $\Phi^{*}_{2n-2}(z)$ in
$\mathbb{C}$ which were also done due to the authors by
Riemann-Hilbert approach in \cite{dd}. The same work of a different
form was also independently given by Mart\'inez-Finkelshtein,
McLaughlin and Saff \cite{mms} at almost the same time.
\end{rem}

\begin{cor}
\begin{equation}
\kappa^{2}_{2n}=\frac{1}{4}[a_{n}^{-2}(1+\beta_{n}^{2})+b_{n}^{-2}].
\end{equation}
\end{cor}

\begin{proof}
Take $z=0$, since $\Phi^{*}_{2n}(0)=1$, then (5.25) follows from
(2.8), (2.10), (2.12), (2.14), (2.15), (2.17) and (5.21).
\end{proof}

\begin{rem}
Recalling the determinant introduced in Remark 3.2, Corollary 5.4
shows that
\begin{equation}
\Lambda_{n}=\begin{vmatrix}
              \lambda_{1,n} & \lambda_{2,n} \\
              -\lambda_{3,n} & -\lambda_{4,n} \\
            \end{vmatrix}=-2\kappa^{2}_{2n}i.
\end{equation}
\end{rem}

As $z\in \mathbb{C}$ is restricted to the unit circle $\partial
\mathbb{D}$, we have

\begin{thm}[OTP and OPUC] Let $\mu$ be a nontrivial probability measure on the
unit circle $\partial \mathbb{D}=\{e^{i\theta}: \theta \in [0,
2\pi)\}$, $\{1, \pi_{n}, \sigma_{n}\}$ be the unique system of the
first orthonormal trigonometric polynomials with respect to $\mu$,
and $\{\Phi_{n}\}$ be the unique system of the monic orthogonal
polynomials on the unit circle with respect to $\mu$. Then for any
$\theta\in [0, 2\pi)$ and $n\in \mathbb{N}$,
\begin{equation}
\Phi_{2n-1}(e^{i\theta})=e^{i(n-1)\theta}[a_{n}\sigma_{n}(\theta)+(\beta_{n}+i)b_{n}\pi_{n}(\theta)]
\end{equation}
and
\begin{equation}
\kappa^{2}_{2n}\Phi^{*}_{2n}(e^{i\theta})=\frac{1}{2}e^{in\theta}[a^{-1}_{n}(1+\beta_{n}i)\sigma_{n}(\theta)
-ib^{-1}_{n}\pi_{n}(\theta)],
\end{equation}
where $\kappa_{2n}$ is the leading coefficient of the orthonormal
polynomial of order $2n$ on the unit circle with respect to $\mu$
and $\kappa_{2n}=\|\Phi_{2n}\|^{-1}_{\mathbb{C}}$, and $a_{n},
b_{n}, \beta_{n}$ are given in (2.17). (5.28) also holds as $n=0$ by
the convention in Section 2.1.
\end{thm}

\begin{cor}
Let $\mu$ be a nontrivial probability measure on the unit circle
$\partial \mathbb{D}=\{e^{i\theta}: \theta \in [0, 2\pi)\}$, $\{1,
\pi_{n}, \sigma_{n}\}$ be the unique system of the first orthonormal
trigonometric polynomials with respect to $\mu$, then
$\pi_{n}(\theta)$ and $\sigma_{n}(\theta)$ have no any common zero
on $[0, 2\pi)$.
\end{cor}

\begin{proof}
It is well known that all zeros of OPUC are located in the open unit
disc. By this fact, Corollary 5.7 follows from (5.27) or (5.28).
\end{proof}

\subsection{Recurrence and Christoffel-Darboux Formulae} Theorem
5.2 and 5.6 give the relationship between orthogonal trigonometric
polynomials and orthogonal polynomials on the unit circle. In view
of the theory of OPUC, we can establish a complete theory for
orthogonal trigonometric polynomials. In this section, as examples,
we give some recurrent formulae and an analogue of
Christoffel-Darboux formula for orthogonal trigonometric
polynomials.

To do so, we introduce reflectional sets, reflectional and
auto-reflectional functions for the unit circle $\partial
\mathbb{D}$.

\begin{defn}
A set $\Sigma$ is called a reflectional set for the unit circle
$\partial \mathbb{D}$ if both $z\in\Sigma$ and
$1/\overline{z}\in\Sigma$, or simply a reflectional set, in which
$z$ and $1/\overline{z}$ are called reflection to  each other. For
example, $\mathbb{C}\setminus \{0\}$ is a reflectional set for the
unit circle.
\end{defn}
\begin{defn}
If $f$ is defined on a reflectional set $\Sigma$, set
\begin{equation}
f_{*}(z)=\overline{f(1/\overline{z})},\,\,z\in \Sigma,
\end{equation}
then $f_{*}$ is the reflectional function of $f$ for the unit
circle, simply reflection.
\end{defn}
\begin{defn}
If $f$ is defined on a reflectional set $\Sigma$ such that
\begin{equation} f(z)=f_{*}(z),\,\,z\in\Sigma,
\end{equation}
then $f$ is called an auto-reflectional function for the unit
circle, simply auto-reflection.
\end{defn}

\begin{prop}
Let $\mu$ be a nontrivial probability measure on the unit circle
$\partial \mathbb{D}=\{z: |z|=1\}$, $\{1, \pi_{n}, \sigma_{n}\}$ be
the unique system of the first orthonormal Laurent polynomials on
the unit circle with respect to $\mu$, then $\pi_{n}$, $\sigma_{n}$
are auto-reflectional in $\mathbb{C}\setminus \{0\}$.
\end{prop}

\begin{proof}
It obviously follows from the fact that $\mu^{(0)}_{n}$ is
auto-reflectional in $\mathbb{C}\setminus \{0\}$, $\mu^{(0)}_{n}$ is
given in (2.8).
\end{proof}

\begin{thm}[Recurrence for Orthogonal Laurent Polynomials]
Let $\mu$ be a nontrivial probability measure on the unit circle
$\partial \mathbb{D}=\{z: |z|=1\}$, $\{1, \pi_{n}, \sigma_{n}\}$ be
the unique system of the first orthonormal Laurent polynomials on
the unit circle with respect to $\mu$. Then for any $z\in
\mathbb{C}\setminus\{0\}$ and $n\in \mathbb{N}\cup\{0\}$,
\begin{align}
&a_{n+1}\sigma_{n+1}(z)+(\beta_{n+1}+i)b_{n+1}\pi_{n+1}(z)\\
=&\frac{1}{2}\kappa^{-2}_{2n}a^{-1}_{n}\big[z(1-\beta_{n}i)-\overline{\alpha}_{2n}(1+\beta_{n}i)\big]\sigma_{n}(z)
+\frac{i}{2}\kappa^{-2}_{2n}b^{-1}_{n}(z+\overline{\alpha}_{2n})\pi_{n}(z)\nonumber
\end{align}
and
\begin{equation}
\begin{cases}
\Re\alpha_{2n-1}=\frac{1}{4}\kappa_{2n}^{-2}[b_{n}^{-2}-a_{n}^{-2}(1-\beta_{n}^{2})],\\[2mm]
\Im\alpha_{2n-1}=-\frac{1}{2}\kappa_{2n}^{-2}a_{n}^{-2}\beta_{n},
\end{cases}
\end{equation}
where $\kappa_{2n}$ is the leading coefficient of the orthonormal
polynomial of order $2n$ on the unit circle with respect to $\mu$,
$\alpha_{2n-1}$, $\alpha_{2n}$ are Verblunsky coefficients, and
$a_{n}$, $b_{n}$, $\beta_{n}$ are given in (2.17). From (5.25) and
(5.32), both $\kappa_{2n-1}$ and $\kappa_{2n}$ can be explicitly
expressed as functions of $a_{n}$, $b_{n}$ and $\beta_{n}$ since
$(1-|\alpha_{2n-1}|)^{1/2}=\kappa_{2n-1}/\kappa_{2n}$, so do all
Verblunsky coefficients $\alpha_{n}$, $n=0,1,\ldots$\,.
\end{thm}

\begin{proof} By Proposition 5.11 and (5.20), we  have
\begin{equation}
\Phi^{*}_{2n-1}(z)=z^{n}[a_{n}\sigma_{n}(z)+(\beta_{n}-i)b_{n}\pi_{n}(z)]
\end{equation}
for $z\in \mathbb{C}$. Since
\begin{equation}
\Phi^{*}_{2n}(z)=\Phi^{*}_{2n-1}(z)-\alpha_{2n-1}z\Phi_{2n-1}(z),
\end{equation}
where $\alpha_{2n-1}$ are Verblunsky coefficients, by (5.20), (5.21)
and (5.33),
\begin{align}
&\left[\frac{1}{2}\kappa^{-2}_{2n}a^{-1}_{n}(1+\beta_{n}i)-(1-\alpha_{2n-1})a_{n}\right]\sigma_{n}(z)\\
=&\left[\frac{1}{2}\kappa^{-2}_{2n}b^{-1}_{n}i+b_{n}\big[(\beta_{n}-i)-\alpha_{2n-1}(\beta_{n}+i)\big]\right]\pi_{n}(z)
\nonumber\end{align} for $z\in \mathbb{C}\setminus\{0\}$. Since
$\sigma_{n}$ and $\pi_{n}$ have no common zeros on the unit circle,
therefore
\begin{equation}
\begin{cases}
\alpha_{2n-1}+\frac{1}{2}a_{n}^{-2}(1+\beta_{n}i)\kappa_{2n}^{-2}=1,\\[2mm]
(\beta_{n}+i)\alpha_{2n-1}-\frac{1}{2}b_{n}^{-2}i\kappa_{2n}^{-2}=\beta_{n}-i.
\end{cases}
\end{equation}
By simple calculations, we obtain
\begin{equation*}
\begin{cases}
\Re\alpha_{2n-1}=\frac{1}{4}\kappa_{2n}^{-2}[b_{n}^{-2}-a_{n}^{-2}(1-\beta_{n}^{2})],\\[2mm]
\Im\alpha_{2n-1}=-\frac{1}{2}\kappa_{2n}^{-2}a_{n}^{-2}\beta_{n}.
\end{cases}
\end{equation*}

Similarly, by (5.20), (5.21) and Proposition 5.11,
\begin{equation}
\Phi_{2n+1}(z)=z^{n}[a_{n+1}\sigma_{n+1}(z)+(\beta_{n+1}+i)b_{n+1}\pi_{n+1}(z)]
\end{equation}
and
\begin{equation}
\Phi_{2n}(z)=\frac{1}{2}\kappa^{-2}_{2n}z^{n}[a^{-1}_{n}(1-\beta_{n}i)\sigma_{n}(z)
+ib^{-1}_{n}\pi_{n}(z)].
\end{equation}
Since
\begin{equation}
\Phi_{2n+1}(z)=z\Phi_{2n}(z)-\overline{\alpha}_{2n}\Phi^{*}_{2n}(z),
\end{equation}
then
\begin{align}
&a_{n+1}\sigma_{n+1}(z)+(\beta_{n+1}+i)b_{n+1}\pi_{n+1}(z)\nonumber\\
=&\frac{1}{2}\kappa^{-2}_{2n}a^{-1}_{n}\big[z(1-\beta_{n}i)-\overline{\alpha}_{2n}(1+\beta_{n}i)\big]\sigma_{n}(z)
+\frac{i}{2}\kappa^{-2}_{2n}b^{-1}_{n}(z+\overline{\alpha}_{2n})\pi_{n}(z)\nonumber
\end{align}
for $z\in \mathbb{C}\setminus\{0\}$.
\end{proof}

\begin{rem}
(5.31) gives a four-terms recurrence for the first orthogonal
Laurent polynomials on the unit circle. Similarly, we can get a
four-terms recurrent formula for the second orthogonal Laurent
polynomials on the unit circle. However, in \cite{dg}, Du and Guo
give a six-terms recurrent formula for the second orthogonal Laurent
polynomials on the unit circle. This difference maybe implies some
identical properties about the coefficients appearing in (5.31) and
the recurrent formula there. (5.32) relates the leading coefficients
and Verblunsky coefficients of OPUC to some coefficients of the
first orthogonal Laurent or trigonometric polynomials on the unit
circle (mainly, $a_{n}$, $b_{n}$ and $\beta_{n}$).
\end{rem}

By the Christoffel-Darboux formula (simply, CD formula) for OPUC and
Theorem 5.2, we have

\begin{thm}[CD formula for Orthogonal Laurent Polynomials]
Let $\mu$ be a nontrivial probability measure on the unit circle
$\partial \mathbb{D}=\{z: |z|=1\}$, $\{1, \pi_{n}, \sigma_{n}\}$ be
the unique system of the first orthonormal Laurent polynomials on
the unit circle with respect to $\mu$. Then for any $z,\zeta\in
\mathbb{C}$ with $\overline{\zeta}z\neq1$ and $n\in
\mathbb{N}\cup\{0\}$,
\begin{equation}
S^{O}_{n}(\zeta, z)+S^{E}_{n}(\zeta,
z)=(1-\overline{\zeta}z)^{-1}D^{O}_{2n+1}(\zeta,z)
\end{equation}
and
\begin{equation}
S^{O}_{n}(\zeta, z)+S^{E}_{n-1}(\zeta,
z)=(1-\overline{\zeta}z)^{-1}D^{E}_{2n}(\zeta,z),
\end{equation}
where
\begin{align}
S^{O}_{n}(\zeta,
z)=&\sum_{j=1}^{n}\kappa_{2j-1}^{2}\overline{\zeta}^{j-1}z^{j-1}\Big[a_{j}^{2}
\overline{\sigma_{j}(\zeta)}\sigma_{j}(z)+b_{j}^{2}(1+\beta_{j}^{2})\overline{\pi_{j}(\zeta)}\pi_{j}(z)\\
&+a_{j}b_{j}\big\{(\beta_{j}+i)\overline{\sigma_{j}(\zeta)}\pi_{j}(z)+
(\beta_{j}-i)\overline{\pi_{j}(\zeta)}\sigma_{j}(z)\big\}\Big],\nonumber
\end{align}

\begin{align}
S^{E}_{n}(\zeta,
z)=&\frac{1}{4}\sum_{j=0}^{n}\kappa_{2j}^{-2}\overline{\zeta}^{j}z^{j}\Big[a_{j}^{-2}(1+\beta_{j}^{2})
\overline{\sigma_{j}(\zeta)}\sigma_{j}(z)+b_{j}^{-2}\overline{\pi_{j}(\zeta)}\pi_{j}(z)\\
&-a_{j}^{-1}b_{j}^{-1}\big\{(\beta_{j}-i)\overline{\sigma_{j}(\zeta)}\pi_{j}(z)+
(\beta_{j}+i)\overline{\pi_{j}(\zeta)}\sigma_{j}(z)\big\}\Big],\nonumber
\end{align}

\begin{equation}
D^{O}_{2n+1}(\zeta,z)=-2i\kappa_{2n+1}^{2}a_{n+1}b_{n+1}\overline{\zeta}^{n}z^{n}
\Big\{\overline{\sigma_{n+1}(\zeta)}\pi_{n+1}(z)-\overline{\pi_{n+1}(\zeta)}\sigma_{n+1}(z)\Big\}
\end{equation}
and
\begin{equation}
D^{E}_{2n}(\zeta,z)=-i\kappa_{2n}^{-2}a^{-1}_{n}b^{-1}_{n}\overline{\zeta}^{n}z^{n}
\Big\{\overline{\sigma_{n}(\zeta)}\pi_{n}(z)-\overline{\pi_{n}(\zeta)}\sigma_{n}(z)\Big\}
\end{equation}
in which $\kappa_{2n+1}$ and $\kappa_{2n}$ only depend on $a_{n}$,
$b_{n}$ and $\beta_{n}$ by (5.25), (5.32) and
$(1-|\alpha_{2n+1}|)^{1/2}=\kappa_{2n+1}/\kappa_{2n+2}$.
\end{thm}

\begin{proof}
By (5.20), (5.21), (5.33) and (5.38), for $z\in \mathbb{C}$
\begin{equation}
\begin{cases}
\varphi_{2n-1}(z)=\kappa_{2n-1}z^{n-1}[a_{n}\sigma_{n}(z)+(\beta_{n}+i)b_{n}\pi_{n}(z)],\\[2mm]
\varphi^{*}_{2n-1}(z)=\kappa_{2n-1}z^{n}[a_{n}\sigma_{n}(z)+(\beta_{n}-i)b_{n}\pi_{n}(z)],
\end{cases}
\end{equation}

\begin{equation}
\begin{cases}
\varphi_{2n}(z)=\frac{1}{2}\kappa^{-1}_{2n}z^{n}[a^{-1}_{n}(1-\beta_{n}i)\sigma_{n}(z)
+ib^{-1}_{n}\pi_{n}(z)],\\[2mm]
\varphi^{*}_{2n}(z)=\frac{1}{2}\kappa^{-1}_{2n}z^{n}[a^{-1}_{n}(1+\beta_{n}i)\sigma_{n}(z)
-ib^{-1}_{n}\pi_{n}(z)].
\end{cases}
\end{equation}

Note that
\begin{align}
S^{O}_{n}(\zeta,
z)=\sum_{j=1}^{n}\overline{\varphi_{2j-1}(\zeta)}\varphi_{2j-1}(z),\,\,\,S^{E}_{n}(\zeta,
z)=\sum_{j=0}^{n}\overline{\varphi_{2j}(\zeta)}\varphi_{2j}(z),
\end{align}

\begin{equation}
D^{O}_{2n+1}(\zeta,z)=\overline{\varphi^{*}_{2n+1}(\zeta)}\varphi^{*}_{2n+1}(z)-
\overline{\varphi_{2n+1}(\zeta)}\varphi_{2n+1}(z)
\end{equation}
and
\begin{equation}
D^{E}_{2n}(\zeta,z)=\overline{\varphi^{*}_{2n}(\zeta)}\varphi^{*}_{2n}(z)-
\overline{\varphi_{2n}(\zeta)}\varphi_{2n}(z).
\end{equation}

Applying the CD formulae (2.31) and (2.32) for OPUC, we get
\begin{equation}
S^{O}_{n}(\zeta, z)+S^{E}_{n}(\zeta,
z)=(1-\overline{\zeta}z)^{-1}D^{O}_{2n+1}(\zeta,z)
\end{equation}
and
\begin{equation}
S^{O}_{n}(\zeta, z)+S^{E}_{n-1}(\zeta,
z)=(1-\overline{\zeta}z)^{-1}D^{E}_{2n}(\zeta,z).\qedhere
\end{equation}
\end{proof}

Restricting $z,\zeta$ to the unit circle, Theorem 5.12 and 5.14
become

\begin{thm}[Recurrence for Orthogonal Trigonometric Polynomials]
Let $\mu$ be a nontrivial probability measure on the unit circle
$\partial \mathbb{D}=\{e^{i\theta}: \theta\in[0, 2\pi)\}$, $\{1,
\pi_{n}, \sigma_{n}\}$ be the unique system of the first orthonormal
trigonometric polynomials with respect to $\mu$. Then for any
$\theta\in[0, 2\pi)$ and $n\in \mathbb{N}\cup\{0\}$,
\begin{align}
&a_{n+1}\sigma_{n+1}(\theta)+(\beta_{n+1}+i)b_{n+1}\pi_{n+1}(\theta)\\
=&\frac{1}{2}\kappa^{-2}_{2n}a^{-1}_{n}\big[e^{i\theta}(1-\beta_{n}i)-\overline{\alpha}_{2n}(1+\beta_{n}i)\big]\sigma_{n}(\theta)
+\frac{i}{2}\kappa^{-2}_{2n}b^{-1}_{n}(e^{i\theta}+\overline{\alpha}_{2n})\pi_{n}(\theta),\nonumber
\end{align}
where $\kappa_{2n}$ is the leading coefficient of the orthonormal
polynomial of order $2n$ on the unit circle with respect to $\mu$,
$\alpha_{2n}$ are Verblunsky coefficients, and $a_{n}$, $b_{n}$,
$\beta_{n}$ are given in (2.17). From (5.25) and (5.32), both
$\kappa_{2n-1}$ and $\kappa_{2n}$ can be explicitly expressed as
functions of $a_{n}$, $b_{n}$ and $\beta_{n}$ since
$(1-|\alpha_{2n-1}|)^{1/2}=\kappa_{2n-1}/\kappa_{2n}$, so do all
Verblunsky coefficients $\alpha_{n}$, $n=0,1,\ldots$\,.
\end{thm}

\begin{thm}[CD formula for Orthogonal Trigonometric Polynomials]
Let $\mu$ be a nontrivial probability measure on the unit circle
$\partial \mathbb{D}=\{e^{i\theta}: \theta\in[0, 2\pi)\}$, $\{1,
\pi_{n}, \sigma_{n}\}$ be the unique system of the first orthonormal
tirgonometric polynomials with respect to $\mu$. Then for any
$\vartheta,\theta\in[0, 2\pi)$ with $\vartheta\neq\theta$ and $n\in
\mathbb{N}\cup\{0\}$,
\begin{equation}
S^{O}_{n}(\vartheta, \theta)+S^{E}_{n}(\vartheta,
\theta)=(1-e^{i(\theta-\vartheta)})^{-1}D^{O}_{2n+1}(\vartheta,\theta)
\end{equation}
and
\begin{equation}
S^{O}_{n}(\vartheta, \theta)+S^{E}_{n-1}(\vartheta,
\theta)=(1-e^{i(\theta-\vartheta)})^{-1}D^{E}_{2n}(\vartheta,\theta),
\end{equation}
where
\begin{align}
S^{O}_{n}(\vartheta,
\theta)=&\sum_{j=1}^{n}\kappa_{2j-1}^{2}e^{i(j-1)(\theta-\vartheta)}\Big[a_{j}^{2}
\sigma_{j}(\vartheta)\sigma_{j}(\theta)+b_{j}^{2}(1+\beta_{j}^{2})\pi_{j}(\vartheta)\pi_{j}(\theta)\\
&+a_{j}b_{j}\big\{(\beta_{j}+i)\sigma_{j}(\vartheta)\pi_{j}(\theta)+
(\beta_{j}-i)\pi_{j}(\vartheta)\sigma_{j}(\theta)\big\}\Big],\nonumber
\end{align}

\begin{align}
S^{E}_{n}(\vartheta,
\theta)=&\frac{1}{4}\sum_{j=0}^{n}\kappa_{2j}^{-2}e^{ij(\theta-\vartheta)}\Big[a_{j}^{-2}(1+\beta_{j}^{2})
\sigma_{j}(\vartheta)\sigma_{j}(\theta)+b_{j}^{-2}\pi_{j}(\vartheta)\pi_{j}(\theta)\\
&-a_{j}^{-1}b_{j}^{-1}\big\{(\beta_{j}-i)\sigma_{j}(\vartheta)\pi_{j}(\theta)
+(\beta_{j}+i)\pi_{j}(\vartheta)\sigma_{j}(\theta)\big\}\Big],\nonumber
\end{align}

\begin{equation}
D^{O}_{2n+1}(\vartheta,\theta)=-2i\kappa_{2n+1}^{2}a_{n+1}b_{n+1}e^{in(\theta-\vartheta)}
\Big\{\sigma_{n+1}(\vartheta)\pi_{n+1}(\theta)-\pi_{n+1}(\vartheta)\sigma_{n+1}(\theta)\Big\}
\end{equation}
and
\begin{equation}
D^{E}_{2n}(\vartheta,\theta)=-i\kappa_{2n}^{-2}a^{-1}_{n}b^{-1}_{n}e^{in(\theta-\vartheta)}
\Big\{\sigma_{n}(\vartheta)\pi_{n}(\theta)-\pi_{n}(\vartheta)\sigma_{n}(\theta)\Big\}
\end{equation}
in which $\kappa_{2n+1}$ and $\kappa_{2n}$ only depend on $a_{n}$,
$b_{n}$ and $\beta_{n}$ by (5.25), (5.32) and
$(1-|\alpha_{2n+1}|)^{1/2}=\kappa_{2n+1}/\kappa_{2n+2}$.
\end{thm}

\subsection{Zeros of Orthogonal Trigonometric Polynomials}

It is well known that all zeros of orthogonal trigonometric
polynomials $\sigma_{n}(\theta)$ and $\pi_{n}(\theta)$ are located
on $[0, 2\pi)$ and they are simple \cite{djy1,dg}. Namely,
$\sigma_n$ and $\pi_n$ possess $2n$ simple zeros in $[0, 2\pi)$, $n$
of which, $\theta_1, \theta_2,\ldots, \theta_n$; $\theta_1^{\prime},
\theta_2^{\prime},\ldots, \theta_n^{\prime}$ are located in $[0,
\pi)$ and the others are just $\pi+\theta_1, \pi+\theta_2, \ldots,
\pi+\theta_n$; $\pi+\theta_1^{\prime}, \pi+\theta_2^{\prime},
\ldots, \pi+\theta_n^{\prime}$. The above Corollary 5.7 shows that
$\theta_{k}\neq\theta^{\prime}_{l}$ for any $k,l=1,2,\ldots,n$.

Moreover, by (5.27) and (5.28), we have

\begin{thm}
Let $\mu$ be a nontrivial probability measure on the unit circle
$\partial \mathbb{D}=\{e^{i\theta}: \theta \in [0, 2\pi)\}$, $\{1,
\pi_{n}, \sigma_{n}\}$ be the unique system of the first orthonormal
trigonometric polynomials with respect to $\mu$, and $\{\Phi_{n}\}$
be the unique system of the monic orthogonal polynomials on the unit
circle with respect to $\mu$, $\theta_1, \theta_2,\ldots, \theta_n$,
$\pi+\theta_1, \pi+\theta_2, \ldots, \pi+\theta_n$ be $2n$ simple
zeros of $\sigma_n$ and $\theta_1^{\prime},
\theta_2^{\prime},\ldots, \theta_n^{\prime}$,
$\pi+\theta_1^{\prime}, \pi+\theta_2^{\prime}, \ldots,
\pi+\theta_n^{\prime}$ be $2n$ simple zeros of $\pi_n$ with
$\theta_j, \theta^{\prime}_j\in[0, \pi)$, $j=1,2,\ldots, n$. Then
for any $1\leq j\leq n$,

\begin{equation}
a_{n}^{-1}=\frac{e^{i(n-1)\theta^{\prime}_j}\sigma_{n}(\theta^{\prime}_j)}{\Phi_{2n-1}(e^{i\theta^{\prime}_j})}=
(-1)^{n-1}\frac{e^{i(n-1)\theta^{\prime}_j}\sigma_{n}(\pi+\theta^{\prime}_j)}{\Phi_{2n-1}(-e^{i\theta^{\prime}_j})},
\end{equation}

\begin{equation}
(\beta_{n}+i)^{-1}b_{n}^{-1}=\frac{e^{i(n-1)\theta_j}\pi_{n}(\theta_j)}{\Phi_{2n-1}(e^{i\theta_j})}=
(-1)^{n-1}\frac{e^{i(n-1)\theta_j}\pi_{n}(\pi+\theta_j)}{\Phi_{2n-1}(-e^{i\theta_j})},
\end{equation}

\begin{equation}
2\kappa^{2}_{2n}a_{n}(1+\beta_{n}i)^{-1}=\frac{e^{in\theta^{\prime}_j}
\sigma_{n}(\theta^{\prime}_j)}{\Phi^{*}_{2n}(e^{i\theta^{\prime}_j})}=
(-1)^{n}\frac{e^{in\theta^{\prime}_j}\sigma_{n}(\pi+\theta^{\prime}_j)}{\Phi^{*}_{2n}(-e^{i\theta^{\prime}_j})}
\end{equation}
and
\begin{equation}
2\kappa^{2}_{2n}b_{n}i=\frac{e^{in\theta_j}\pi_{n}(\theta_j)}{\Phi^{*}_{2n}(e^{i\theta_j})}=
(-1)^{n}\frac{e^{in\theta_j}\pi_{n}(\pi+\theta_j)}{\Phi^{*}_{2n}(-e^{i\theta_j})},
\end{equation}
where $\kappa_{2n}$ is the leading coefficient of the orthonormal
polynomial of order $2n$ on the unit circle with respect to $\mu$
and $\kappa_{2n}=\|\Phi_{2n}\|^{-1}_{\mathbb{C}}$, and $a_{n},
b_{n}, \beta_{n}$ are given in (2.17).
\end{thm}

\begin{proof}
It easily follows from (5.27) and (5.28) since $\sigma_{n}$ and
$\pi_{n}$ have no common zeros by Corollary 5.7.
\end{proof}

\begin{rem}
In fact, in terms of the zeros of orthogonal trigonometric
polynomials, the above theorem gives the solutions of four classes
of equations as follows
\begin{equation}
e^{i(n-1)x}\sigma_{n}(x)-a_{n}^{-1}\Phi_{2n-1}(e^{ix})=0,
\end{equation}
\begin{equation}
e^{inx}\sigma_{n}(x)-2\kappa^{2}_{2n}a_{n}(1+\beta_{n}i)^{-1}\Phi^{*}_{2n}(e^{ix})=0,
\end{equation}
\begin{equation}
e^{i(n-1)x}\pi_{n}(x)-(\beta_{n}+i)^{-1}b_{n}^{-1}\Phi_{2n-1}(e^{ix})=0
\end{equation}
and
\begin{equation}
e^{inx}\pi_{n}(x)-2\kappa^{2}_{2n}b_{n}i\Phi^{*}_{2n}(e^{ix})=0,
\end{equation}
where $x\in[0,2\pi)$.
\end{rem}

Noting the determinant $\Lambda_{n}$ introduced in Remark 3.2, by
Theorem 5.17, we can get much more properties of zeros. For example,
we have

\begin{thm}
Let $\mu$ be a nontrivial probability measure on the unit circle
$\partial \mathbb{D}=\{e^{i\theta}: \theta \in [0, 2\pi)\}$, $\{1,
\pi_{n}, \sigma_{n}\}$ be the unique system of the first orthonormal
trigonometric polynomials with respect to $\mu$, and $\{\Phi_{n}\}$
be the unique system of the monic orthogonal polynomials on the unit
circle with respect to $\mu$, $\theta_1, \theta_2,\ldots, \theta_n$,
$\pi+\theta_1, \pi+\theta_2, \ldots, \pi+\theta_n$ be $2n$ simple
zeros of $\sigma_n$ and $\theta_1^{\prime},
\theta_2^{\prime},\ldots, \theta_n^{\prime}$,
$\pi+\theta_1^{\prime}, \pi+\theta_2^{\prime}, \ldots,
\pi+\theta_n^{\prime}$ be $2n$ simple zeros of $\pi_n$ with
$\theta_j, \theta^{\prime}_j\in[0, \pi)$, $j=1,2,\ldots, n$. Then

\begin{equation}
\begin{vmatrix}
  \displaystyle\frac{\Phi_{2n-1}(e^{i\theta_{k}^{\prime}})}{e^{i(n-1)\theta_{k}^{\prime}}\sigma_{n}(e^{i\theta_{k}^{\prime}})}
   & \displaystyle\frac{\Phi_{2n-1}(e^{i\theta_{l}})}{e^{i(n-1)\theta_{l}}\pi_{n}(e^{i\theta_{l}})}
   \\[6mm]
  \displaystyle\frac{\Phi^{*}_{2n}(e^{i\theta_{s}^{\prime}})}{e^{in\theta_{s}^{\prime}}\sigma_{n}(e^{i\theta_{s}^{\prime}})}
  & \displaystyle\frac{\Phi^{*}_{2n}(e^{i\theta_{t}})}{e^{in\theta_{t}}\pi_{n}(e^{i\theta_{t}})} \\
\end{vmatrix}=-2a_{n}b_{n}i,
\end{equation}
for any $1\leq k,l,s,t\leq n$, where $a_{n}, b_{n}$ are given in
(2.17).
\end{thm}

\begin{proof}
Note that
\begin{equation}
a_{n}b_{n}\Lambda_{n}=\begin{vmatrix}
                        a_{n} & (\beta_{n}+i)b_{n} \\[3mm]
                        \frac{1}{2}a_{n}^{-1}(1+\beta_{n}i) & -\frac{1}{2}b_{n}^{-1}i \\
                      \end{vmatrix},
\end{equation}
then (5.68) follows from Remark 5.5 and Theorem 5.17.
\end{proof}

In the conclusion of the present paper, we again use the determinant
$\Lambda_{n}$ to express the first orthogonal Laurent polynomials on
the unit circle in terms of orthogonal polynomials on the unit
circle.

To do so, let
\begin{equation*}
\Lambda(n)=\begin{pmatrix}
             1 & \beta_{n}+i \\[3mm]
             \frac{1}{2}a_{n}^{-2}(1+\beta_{n}i) & -\frac{1}{2}b_{n}^{-2}i \\
           \end{pmatrix},
\end{equation*}

\begin{equation*}
T_{n}(z)=\begin{pmatrix}
           a_{n}\sigma_{n}(z) \\[3mm]
           b_{n}\pi_{n}(z) \\
         \end{pmatrix}
\end{equation*}
 and
\begin{equation*}
U_{n}(z)=\begin{pmatrix}
           z^{-n+1}\Phi_{2n-1}(z) \\[3mm]
           \kappa^{2}_{2n}z^{-n}\Phi^{*}_{2n}(z) \\
         \end{pmatrix}.
\end{equation*}
$\Lambda(n)$ is called connection matrix. Then
$\Lambda_{n}=\det\big(\Lambda(n)\big)$ and
\begin{equation*}
U_{n}(z)=\Lambda(n)T_{n}(z)
\end{equation*}
holds for $z\in \mathbb{C}\setminus\{0\}$ and $n\in \mathbb{N}$.
Since $\Lambda_{n}\neq 0$ then $\Lambda(n)$ is invertible. Hence,
\begin{equation*}
T_{n}(z)=\Lambda^{-1}(n)U_{n}(z)
\end{equation*}
for $z\in \mathbb{C}\setminus\{0\}$ and $n\in \mathbb{N}$, where
$\Lambda^{-1}(n)$ is the inverse of $\Lambda(n)$. Explicitly,
\begin{equation*}
\begin{cases}
a_{n}\sigma_{n}(z)=\frac{1}{2}z^{-n}[2\Lambda_{n}^{-1}z\Phi_{2n-1}(z)+(1-\beta_{n}i)\Phi^{*}_{2n}(z)] \\[3mm]
b_{n}\pi_{n}(z)=-\frac{1}{4}z^{-n}[2\Lambda_{n}^{-1}a^{-2}_{n}(1+\beta_{n}i)z\Phi_{2n-1}(z)-b_{n}^{-2}\Phi^{*}_{2n}(z)]
\end{cases}
\end{equation*}
for $z\in \mathbb{C}\setminus\{0\}$ and $n\in \mathbb{N}$, where
$\Lambda_{n}=-\frac{1}{2}[a_{n}^{-2}(1+\beta_{n}^{2})+b_{n}^{-2}]i$,
and $a_{n}, b_{n}, \beta_{n}$ are given in (2.17).

% ----------------------------------------------------------------
%\bibliographystyle{amsplain}
%\bibliography{6}

\end{document}